\documentclass[twocolumn,preprintnumbers,amsmath,amssymb]{revtex4-1}

\usepackage{graphicx}
\usepackage{amssymb}
\usepackage{hyperref}
\usepackage{color}

\begin{document}

\title{Single crystal study of the charge density wave metal LuNiC$_2$}

\author{S. Steiner} 
\author{H. Michor}\email{michor@ifp.tuwien.ac.at} 
\author{O. Sologub} 
\author{B. Hinterleitner}
\author{F. H{\"o}fenstock} 
\author{M. Waas} 
\author{E. Bauer} 
\affiliation{Institute of Solid State Physics, TU Wien, A-1040 Wien, Austria}

\author{B. St{\"o}ger} 
\affiliation{X-Ray Center, TU Wien, Getreidemarkt 9, A-1060 Wien, Austria}

\author{V. Babizhetskyy} 
\author{V. Levytskyy} 
\author{B. Kotur}

\affiliation{Department of Inorganic Chemistry, Ivan Franko National University of Lviv, Kyryla and Mefodiya Str., 6, UA-79005 Lviv, Ukraine}

\begin{abstract}
We report on single crystal growth, single crystal x-ray diffraction, physical properties and density functional theory (DFT) 
electronic structure as well as Fermi surface calculations for two ternary carbides, LuCoC$_2$ and LuNiC$_2$. 
Electrical resistivity measurements reveal for LuNiC$_2$ a charge density wave (CDW) transition at $T_{\rm CDW}\simeq 450$\,K
and, for $T>T_{\rm CDW}$, a significant anisotropy of the electrical resistivity, which is lowest along the orthorhombic $a$-axis. 
The analysis of x-ray superstructure reflections suggest a commensurate CDW state with a Peierls-type distortion of the Ni atom 
periodicity along the orthorhombic $a$-axis. 
DFT calculations based on the CDW modulated monoclinic structure model of LuNiC$_2$ as compared to 
results of the orthorhombic parent-type reveal the formation of a partial CDW gap at the Fermi level
which reduces the electronic density of states from $N(E_{\rm F})=1.03$ states/eV\,f.u.\ without CDW to 
$N(E_{\rm F})=0.46$ states/eV\,f.u.\ in the CDW state.
The corresponding bare DFT Sommerfeld value of the latter, $\gamma_{\rm DFT}^{\rm CDW}=0.90$\,mJ/mol\,K$^2$,
reaches reasonable agreement with the experimental value $\gamma=0.83(5)$\,mJ/mol\,K$^2$ of LuNiC$_2$.
LuCoC$_2$ displays a simple metallic behavior with neither CDW ordering nor superconductivity above 0.4\,K. 
Its experimental Sommerfeld coefficient, $\gamma=5.9$(1)\,mJ/mol\,K$^2$, is in realistic correspondence with the calculated, 
bare Sommerfeld coefficient, $\gamma_{{\rm DFT}}=3.82$\,mJ/mol\,K$^2$, of orthorhombic LuCoC$_2$. 
\end{abstract}


\date{\today}

\maketitle

\section{Introduction}

Intermetallic rare earth nickel dicarbides, $R$NiC$_2$ ($R=$ La, \dots\ Lu), with the 
non-centrosymmetric orthorhombic CeNiC$_2$ structure-type~\cite{bodak80,jeitschko}, 
exhibit a variety of exciting physical phenomena. The initial interest focused i) 
on their rare earth magnetism~(see e.g.\ Refs.~\cite{Kotsanidis89,schaefer97,OnoderaKoshikawa98}) and, 
subsequently, ii) on LaNiC$_2$, which exhibits superconductivity below about $T_c=2.9$\,K~\cite{lee,HiroseKishino12} 
with a time reversal symmetry broken order parameter~\cite{Hillier,Hase}, and 
iii) on the multiple charge density wave (CDW) transitions of PrNiC$_2$, NdNiC$_2$, \dots,  TmNiC$_2$
\cite{murase,shimomura,PhysRevB.97.041103} and, finally, iv) on the complex
interplay of magnetic and incommensurate as well as commensurate CDW order 
parameters~\cite{KimRhyee12,Prathiba2016,kolincio16,LeiWang17,kolincio17}  
(see Ref.~\cite{BABIZHETSKYY2017} for a review).
The Peierls temperature, i.e.\ onset of CDW order, increases inversely proportional to the unit cell 
volume of $R$NiC$_2$ compounds~\cite{PhysRevB.97.041103} and is, thus, largest for LuNiC$_2$, which 
has not yet been studied in closer detail. 

In the present work we investigate two lutetium $3d$-metal dicarbides,
LuCoC$_2$ and LuNiC$_2$ (initially reported by Jeitschko~\cite{jeitschko}), with respect to
their crystal structure, their electronic ground state properties, and in particular, with respect to the occurrence of 
charge density wave or superconducting transitions, by means of specific heat, magnetization, and 
electrical resistivity measurements as well as computational electronic structure and Fermi surface studies.

\section{Experimental details}
\label{ED}

Polycrystalline samples, LuCoC$_2$ and LuNiC$_2$, have been prepared by arc melting with subsequent annealing 
at 1000\,${^\circ}$C for 10 days using a preparation procedure described earlier~\cite{michor15}. 
Commercially available elements, Lu distilled bulk pieces (Metall Rare Earth, purity of 99.9 at.\%
and 99.99\,\%\ Lu/$R$), powders of electrolytic nickel and cobalt (Strem Chemicals, 99.99 at.\%), 
and graphite powder (Aldrich, 99.98 at.\%) were used. 

A first attempt to grow LuCoC$_2$ and LuNiC$_2$ single crystals has been conducted via the Czochralski method, by 
which we have earlier succeeded to grow HoCoC$_2$~\cite{michor17}. 
In the present work on LuNiC$_2$, however, Czochralski pulling resulted in 
rather small single crystalline domains (of the order of cubic-mm size) and just slightly larger crystalline domains 
in the case of LuCoC$_2$. The latter has been used for a specific heat measurement (see below). 

A large single crystal of LuNiC$_2$ ($> 200$ mm$^3$) was finally grown from a stoichiometric polycrystalline feed 
rod via the floating zone technique in an optical mirror furnace (Crystal Systems Corporation, Japan) and 
was oriented by means of the Laue method.
A cross-section of the crystal (parallel and perpendicular to the growth direction) was examined with a scanning electron
microscope (SEM) using a Philips XL30 ESEM with EDAX XL-30 EDX-detector. 
Thereby, a relatively homogeneous distribution of small inclusions (multi-phase precipitates of typically 5\,--\,10 $\mu$m size, 
essentially composed from transparent Lu-C-O crystals and metallic Lu-Ta-C as well as Ni rich Lu-Ni-C phases) 
inside the single crystalline matrix LuNiC$_2$ is resolved by SEM, 
but remains below the resolution limit of the powder X-ray diffraction (XRD) pattern (see below).  
From the density of precipitates in the SEM image, we roughly estimate a volume fraction of impurity phases 
of the order of 0.1\,\%\ of the total crystal volume.

Powder XRD data of LuCoC$_2$ and LuNiC$_2$ were collected on a Siemens D5000 powder diffractometer 
with graphite monochromated Cu-K$_{\alpha}$ radiation (20$^{\circ} \leq 2\Theta \leq 120^{\circ}$, 
step size 0.02$^{\circ}$). 
While powder XRD revealed the presence of some impurity phases in the polycrystalline material
of LuCoC$_2$ and LuNiC$_2$, Bragg intensities due to impurity phases have neither 
been resolved in powder XRD for the Czochralski grown crystalline material of LuCoC$_2$ and LuNiC$_2$, 
nor for pieces of the zone refined LuNiC$_2$ crystal.

Crystals for XRD were isolated via mechanical fragmentation of the annealed LuNiC$_2$ and LuCoC$_2$ samples. 
Single crystal X-ray intensity data were collected at $T = 100$\,K and 298\,K on a four-circle
Bruker APEX II diffractometer (CCD detector, $\kappa$-geometry, Mo K$_{\alpha}$-radiation, $\lambda = 0.71073$~\AA).
Multi-scan absorption correction was applied using the program \texttt{SADABS}; frame data were reduced to 
intensity values applying the \texttt{SAINT-Plus} package~\cite{Bruker}. 
The structures were solved by direct methods and refined with the \texttt{SHELXS-97} and \texttt{SHELXL-97} 
programs \cite{SHELXS,SHELXL}, respectively. 

Zero-field specific heat data of single-crystalline samples of about 60~mg of LuCoC$_2$ and 100~mg of LuNiC$_2$
were collected in the temperature range 400\,mK to 15\,K using a PPMS $^3$He heat capacity insert.
 
Four probe resistivity measurements were carried out on bar shaped samples with
contacts made by spot welding thin gold wires ($d = 50$\,$\mu$m) on the sample surface. 
For better mechanical stability, the spot welded contacts were coated with silver epoxy. 
LuNiC$_2$ single crystalline bar shaped samples were cut along the
principle crystallographic orientations with typical dimensions $0.7\times 0.9\times 5$~mm$^3$ and measured 
in a PPMS resistivity set-up (2\,--\,400\,K) as well as in a home-made furnace set up (300\,--\,610\,K) 
with the identical set of samples. 
High temperature resistivity measurements of polycrystalline samples LuCoC$_2$ and LuNiC$_2$ were performed in an 
ULVAC ZEM-3 measuring system (300\,K\,--\,650\,K).
Temperature and field dependent magnetic measurements were carried out on a CRYOGENIC SQUID magnetometer in 
a temperature range from 2\,K to room temperature applying static magnetic fields up to 7\,T.

\section{Computational methods}

\begin{figure}[b] 
\begin{center}
\includegraphics[width=0.65\columnwidth]{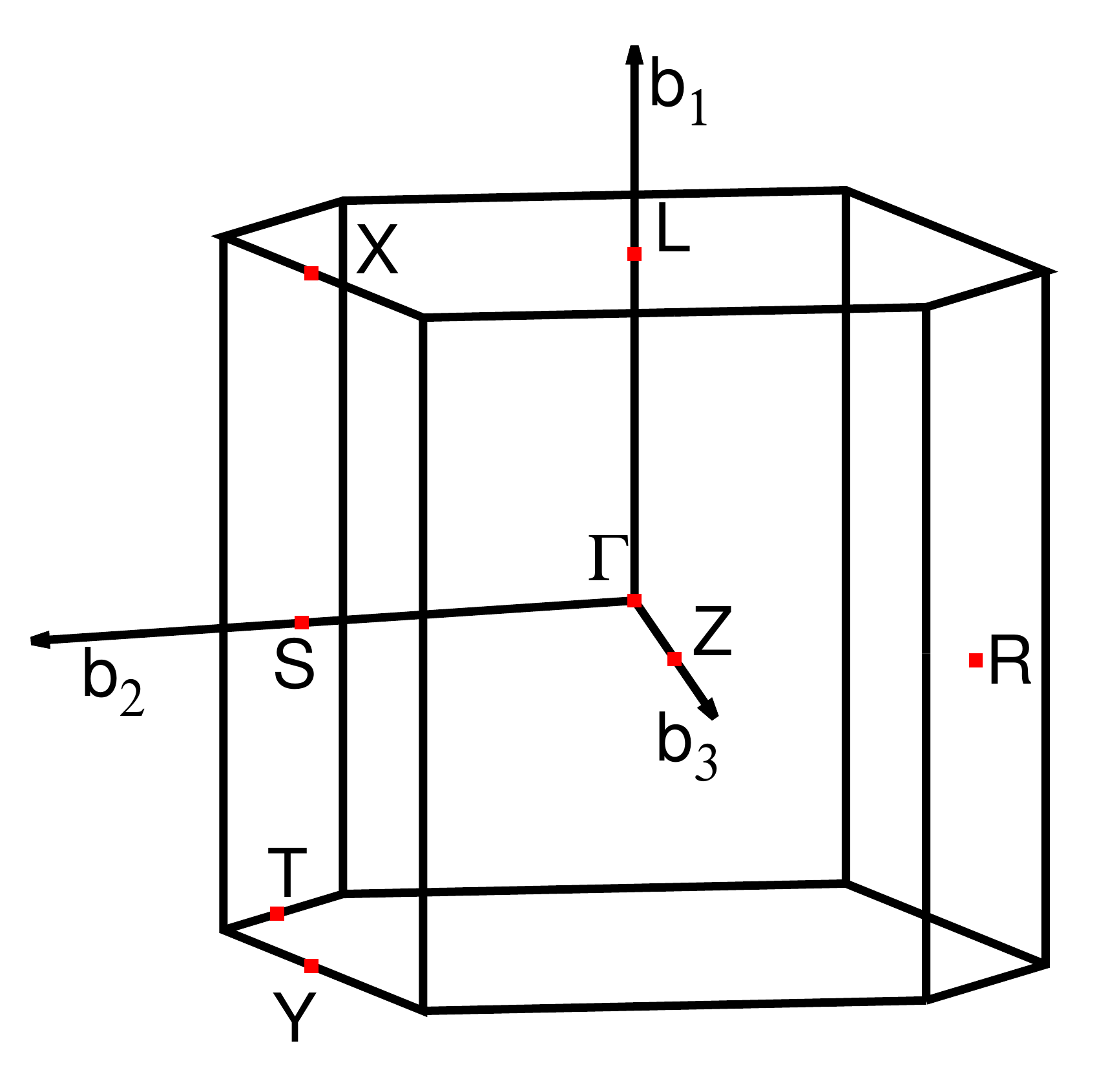}
\caption{The first Brillouin zone of the orthorhombic CeNiC$_2$-type structure including high-symmetry points (see text).}
\label{BZ}
\end{center}
\end{figure}

Density functional theory (DFT) calculations based on the CeNiC$_2$-type orthorhombic structure models of LuCoC$_2$ 
as well as LuNiC$_2$ were performed within the projector augmented wave methodology
\cite{bloechl1994} implemented in the \texttt{Vienna ab initio simulation package} (VASP)
\cite{kresse1996, kresse1999}. The generalized gradient approximation, as parametrized by Perdew, Burke, 
and Ernzerhof \cite{perdew1996}, was applied for the exchange potential.
For the valence state configurations of the pseudo-potentials we included the $5$d$^{1}6$p$^{6}6$s$^{2}$ states for Lu 
(the $4$f-electrons were kept frozen in the core), the $3$p$^{6}3$d$^{8}4$s$^{1}$ states for Co, the $3$p$^{6}3$d$^{9}4$s$^{1}$ 
states for Ni, and the $2$s$^{2}$p$^{2}$ states for the C-atoms. The remaining electrons were kept frozen.

The atom positions were relaxed with fixed experimental lattice constants (see Table~\ref{tab1}).
During the relaxation, the total energy was minimized until the energy convergence became better than $1 \times 10^{-8}$ eV.
To obtain the equilibrium positions of the ions, residual forces were optimized until they under-matched
$1 \times 10^{-2}$ eV/\AA.
The Brillouin zone integration for the relaxation,  was done on  a $10 \times 8 \times 6$ grid of $\mathbf{k}$-points on a 
Monkhorst and Pack \cite{monkhorst1976} mesh using the method of Methfessel-Paxton \cite{methfessel1989} of the order one
with a smearing parameter $\sigma=0.1$, which results in well-converged total energies and optimized ionic positions. 
We note that complementary DFT calculations with a full relaxation of the CeNiC$_2$-type structure model (i.e.\ including 
the volume and lattice constants) closely reproduced all the below results obtained with a fixed cell shape. 
 
The electronic band structure (BS) was calculated using fully relaxed ionic positions with fixed cell shape and volume
taking into account spin orbit coupling (SOC) for the valence electrons as described in~\cite{ssteiner2016}. 
For the evaluation of the electronic density of states (EDOS) a dense $20 \times 16 \times 12$ $\mathbf{k}$-point mesh
has been implemented. The $\mathbf{k}$-point integration was done using the tetrahedron method with Bl\"{o}chl corrections
\cite{bloechl1994}.
The Brillouin zone integration for the BS was done using again the method of Methfessel-Paxton \cite{methfessel1989}.
The electronic BS is plotted along high-symmetry $\mathbf{k}$-points in the irreducible Brillouin zone (IRBZ) as depicted in Fig.~\ref{BZ}.
These special $\mathbf{k}$-points in reciprocal space are: $\Gamma=(0,0,0)$, $S=(0,\frac{1}{2},0)$,
$R=(0,-\frac{1}{2},\frac{1}{2})$, $Z=(0,0,\frac{1}{2})$, $Y=(-\frac{1}{2},\frac{1}{2}, 0)$,
$T=(-\frac{1}{2},\frac{1}{2},-\frac{1}{2})$, $X=(\frac{1}{2},\frac{1}{2},0)$ and $L=(\frac{1}{2},0,0)$,
these $\mathbf{k}$-points are either a center of a face in IRBZ,
a corner of IRBZ or a midpoint of a line-edge. 
We note that switching the space group setting from $Amm2$ to $Cm2m$, e.g.\ used in Refs.~\cite{IzumiHase09,HiroseKishino12}, leads to 
a commutation of axis $(a,b,c)$ to $(c,a,b)$ and, thus, to correspondent changes of the labeling of special 
$\mathbf{k}$-points in the IRBZ.

The Fermi surface is interpolated using maximally localized Wannier functions as implemented
in the \texttt{WANNIER90} code~\cite{mostofi2008}. 
We have ascertained that the interpolated Wannier bands match well
with the VASP bands in the region of interest, thus, being suited to obtain reasonable Fermi 
surfaces.

For LuNiC$_2$, additional DFT calculations of the EDOS were performed for the CDW modulated structure 
model presented in section~\ref{CS_XRD} (see Table~\ref{tab2})
with a slightly adapted procedure: a variation of the volume and the cell shape was included in the 
relaxation procedure to overcome convergence problems, though the relative changes of the lattice constants during 
their relaxation did not exceed 3\,\%\ and the relative deviation from the experimental volume remained
within $2\times 10^{-3}$. Relaxed atom coordinates are reasonably close to the experimental ones, e.g.\ 
yielding alternating Ni--Ni distances along the monoclinic $b$-axis of 3.247~\AA{} and 3.665~\AA{} 
as compared to the experimentally refined values 
of 3.208(2)~\AA{} and 3.682(2)~\AA{} (see below). 
The optimization of residual forces was conducted with the same criteria, $1 \times 10^{-2}$ eV/\AA, as given above. 
For the Brillouin zone integration steps during relaxation, 
a grid of $5 \times 5 \times 9$ $\mathbf{k}$-points on a Monkhorst and Pack \cite{monkhorst1976} mesh has been employed 
using the same method as for the CeNiC$_{2}$-type structure model.

The electronic DOS of the CDW modulated structure was calculated using the fully relaxed structure including SOC
on a $\mathbf{k}$-point mesh of $9 \times 10 \times 18$ with the tetrahedron method with Bl\"{o}chl corrections
\cite{bloechl1994} and $\sigma=0.05$. The $\mathbf{k}$-point mesh used for the CDW-superstructure is lower than the mesh
used for CeNiC$_{2}$-type structure, because of its two times larger volume in real and, hence, 
correspondingly smaller Brillouin zone volume in the reciprocal space. Fermi surface calculations were performed
with the same methods as for the parent-type structure model.

\section{Results}

\subsection{Crystal Structure Determination from Single Crystal XRD Data}
\label{CS_XRD}

Single crystal XRD data for LuNiC$_2$ and LuCoC$_2$ indicate a high degree of order (mosaicities $<0.50$).
In contrast to LuCoC$_2$, LuNiC$_2$ features distinct superstructure reflections.
Disregarding these reflections, the diffraction patterns of both, LuNiC$_2$ and LuCoC$_2$, evidence 
orthorhombic symmetry with similar cell parameters.
Systematic absences were consistent with the space groups, $A222$, $A2mm$, $Am2m$, $Amm2$, and $Ammm$, 
out of which the non-centrosymmetric $Amm2$ proved to be correct during structure solution and refinement,
as proposed earlier in Ref.~\cite{jeitschko}. 
Lu and transition metal atom positions were deduced from direct methods with \texttt{SHELXS-97} and refined in 
a straightforward manner using \texttt{SHELXS-97}; carbon sites were easily located in the difference Fourier map. 
The refinements with free site occupation factors revealed some indication that the cobalt position in 
LuCoC$_2$ is occupied by about 98.4(6)\%, whereas no deviations from full occupancies 
have been detected for Lu and Ni atom positions. 
The final positional and atom displacement parameters obtained from single crystals are listed 
in Table~\ref{tab1}. A significant anisotropy of the atom displacement parameter with a substantially increased $U_{11}$
($U_{11}\simeq 3.6\times U_{22}\simeq 3.2\times U_{33}$) is observed only for Ni in LuNiC$_2$
which manifests some trace of CDW order in the refinement based on the orthorhombic CeNiC$_2$-type structure model. 

\begin{table}[b]
	\caption{Unit cell parameters (structure type CeNiC$_2$, space group $Amm2$, no. 38, $Z=2$), atomic coordinates 
	and anisotropic displacement parameters as well as experimental and refinement parameters; 
	crystal structure data are standardized using the program Structure Tidy~\protect\cite{Typix}.
	Diffraction data were collected at $T=298$\,K.}
	\begin{center}
	\renewcommand{\arraystretch}{1.4}
		\begin{tabular}{lll}
			\hline\hline
			   & LuCoC$_2$ & LuNiC$_2$ \\
			\hline
		$a$ (\AA) & 3.4226(2) &	3.4506(2)  \\
	  $b$ (\AA) & 4.4895(3) &	4.4787(2)  \\
		$c$ (\AA) & 5.9916(4)	& 5.9857(3)  \\
			\hline
	  Lu in $2b$ $(\frac{1}{2},0,z)$ & $z=0.3835(1)$ &	$z=0.3885(2)$  \\
		occupation factor  & 1.00 & 1.00 \\
	  $U_{11}^b$ & 0.0079(1)  &	0.0070(1) \\
		$U_{22}$ & 0.0080(1) &	0.0108(1)  \\
		$U_{33}$ & 0.0085(1) &	 0.0084(1)  \\
		\hline
		 $M$ in $2a$ $(0,0,0)$ & &	  \\
		occupation factor   & 0.984\,(6) & 1.00 \\
	  $U_{11}^b$ & 0.0098(4)   &	0.0217(5) \\
		$U_{22}$ & 0.0060(3) &	0.0060(3)  \\
		$U_{33}$ & 0.0070(4) &	 0.0068(4)  \\
		\hline
		 C in $4d$  $(0,y,z)$ & $y=0.3458(9)$ &	$y=0.3444(11)$ \\
	                     & $z=0.1890(7)$ &	$z=0.1873(9)$  \\
   occupation factor   & 1.00 & 1.00 \\
	$U_{iso}$ & 0.0088(6) &	0.0097(7)  \\
				\hline
			\hline
			Theta range (deg) & $5.68<\theta <37.05$	& $5.69<\theta <34.87$ \\
	  Crystal size ($\mu$m) & $48\times 50\times 55$	& $55\times 55\times 60$ \\
		Reliability factors$^c$ & $R_F^2=0.0095$ & $R_F^2=0.0119$ \\
		GOF & 1.043 & 1.166 \\
		Extinction  & 0.0149(9) &	0.0144(10) \\
		(Zachariasen) &  &	 \\
		Residual density:   & 0.801; -0.742 &	0.847; -1.362 \\
		max; min (e$^{-}$/\AA$^3$) & & \\
			\hline
		\multicolumn{3}{l}{$^a$ $M =$ Co or Ni; } \\
		\multicolumn{3}{l}{$^b$  $U_{23}=U_{13}=U_{12}=0$;} \\
		\multicolumn{3}{l}{$^c$  $R_F^2=\sum{\left| F_0^2-F_c^2\right|}/\sum{F_0^2}$} \\
		\hline\hline
		\end{tabular}	
		\end{center}
	\label{tab1}
\end{table}

\begin{table}[b]
    \caption{Unit cell parameters and atomic coordinates
    as well as experimental and refinement parameters of 
		the CDW superstructure of LuNiC$_2$
		(space group $Cm$, no. 8) measured at $T=100$\,K.}
    \begin{center}
    \renewcommand{\arraystretch}{1.4}
        \begin{tabular}{ll}
            \hline\hline
               & LuNiC$_2$ \\
            \hline
        $a$ (\AA) & 7.4662(9) \\
        $b$ (\AA) & 6.8897(9) \\
        $c$ (\AA) & 3.7326(5) \\
        $\beta$ ($^\circ$) & 106.324(3) \\
            \hline
    Lu1 in $2a$ $(x,\frac1 2,z)$ & $x=-0.00866(6)$  \\
                 & $z=-0.00807(8)$  \\
    Lu2 in $2a$ $(x,0,z)$ & $x=0.00866(5)$  \\
                 & $z=0.00807(8)$  \\
    Ni in $4b$ & $x=0.1929(9)$  \\
                 & $y=0.2672(2)$  \\
                 & $z=0.6087(16)$  \\
    C1 in $4b$ & $x=0.425(4)$ \\
                 & $y=0.2577(8)$  \\
                 & $z=0.448(6)$  \\
    C2 in $4b$ & $x=0.268(4)$ \\
                 & $y=0.2501(7)$  \\
                 & $z=0.146(7)$  \\
                \hline
            \hline
            Theta range (deg) & $4.13<\theta <49.68$ \\
      Crystal size ($\mu$m) & $80\times 60\times 20$ \\
        Reliability factors  & $R_F^2=0.0294$ \\
        ($R_F^2=\sum{\left| F_0^2-F_c^2\right|}/\sum{F_0^2}$) & \\
        GOF & 1.096 \\
        Extinction  &    0.0107(12) \\
        (Zachariasen)   &     \\
        Residual density:   &     3.097; $-3.652$ \\
        max; min (e$^{-}$/\AA$^3$)  & \\
        \hline\hline
        \end{tabular}
        \end{center}
    \label{tab2}
\end{table}

\begin{figure}[t]
	\begin{center}
	\includegraphics[width=0.49\columnwidth]{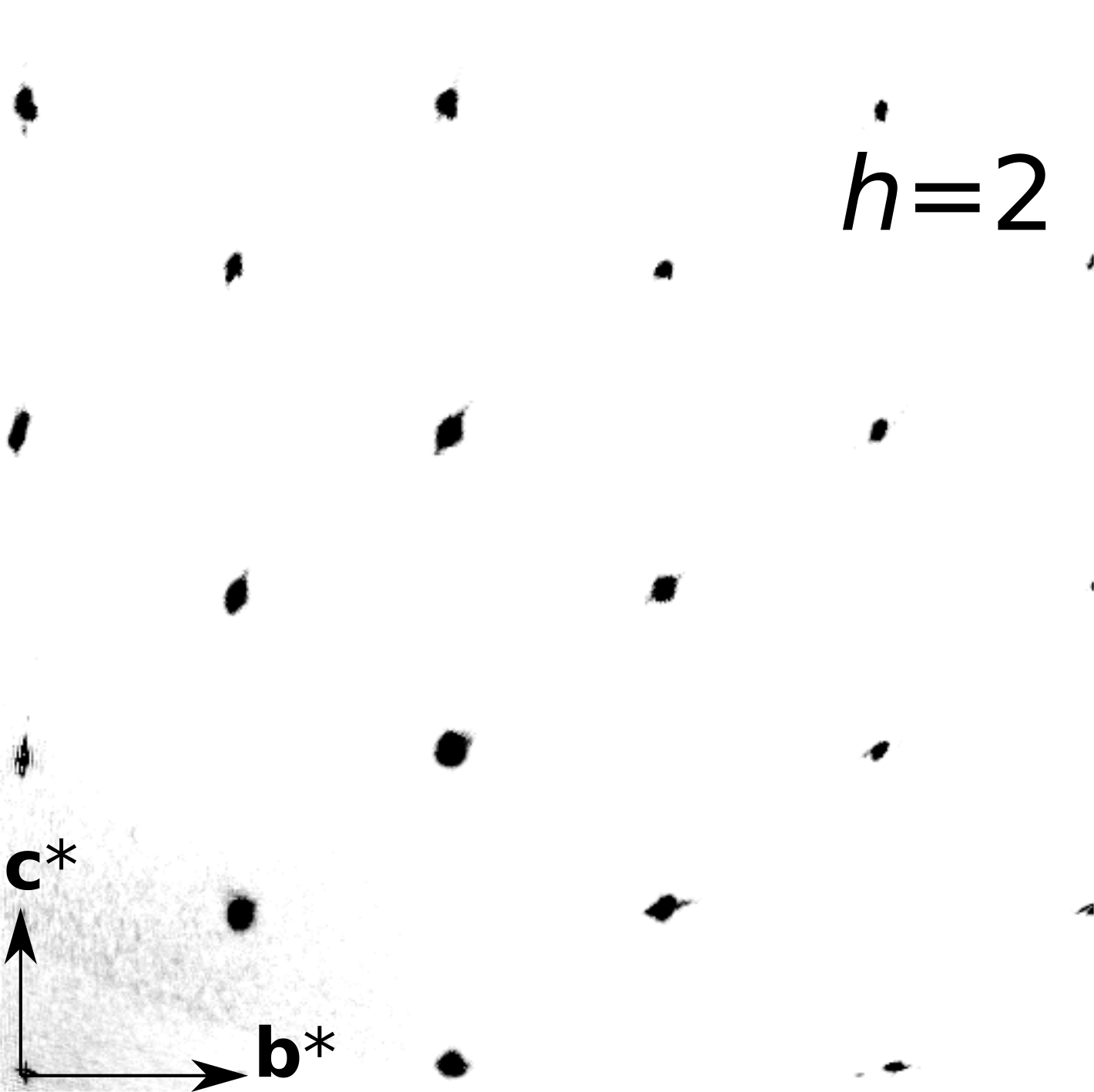}
	\includegraphics[width=0.49\columnwidth]{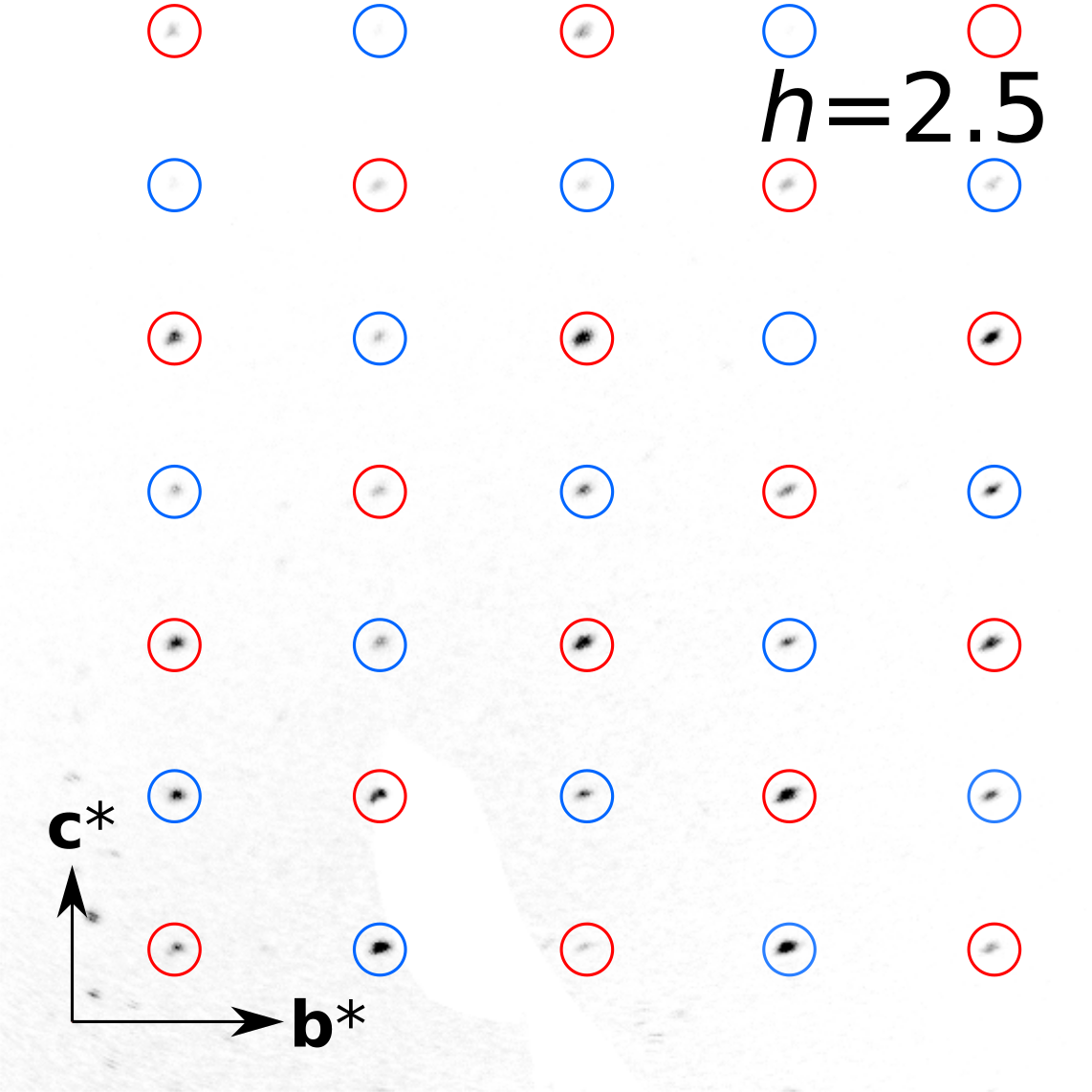}
			\caption{$h=2$ and $h=2.5$ planes with respect to the reciprocal lattice of the 
			CeNiC$_2$-type parent structure of LuNiC$_2$ reconstructed from CCD data in the 
			left and the right panel, respectively. 
		Super-structure reflections attributed to two different twin domains
		are marked by red and blue circles.}
	\label{reci}
	\end{center}
\end{figure}

\begin{figure}[t]
	\begin{center}
	\includegraphics[width=0.80\columnwidth]{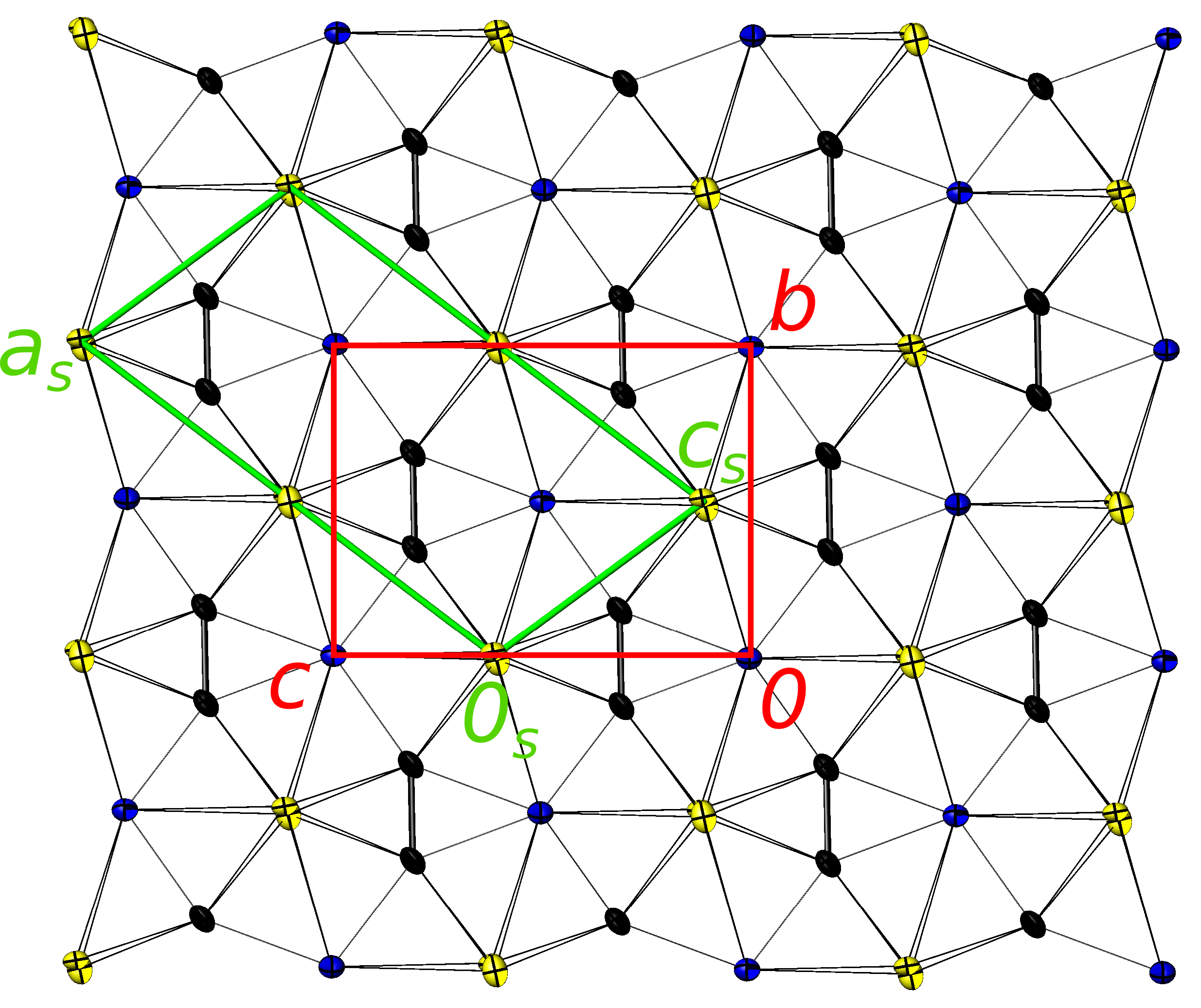}
	\includegraphics[width=0.70\columnwidth]{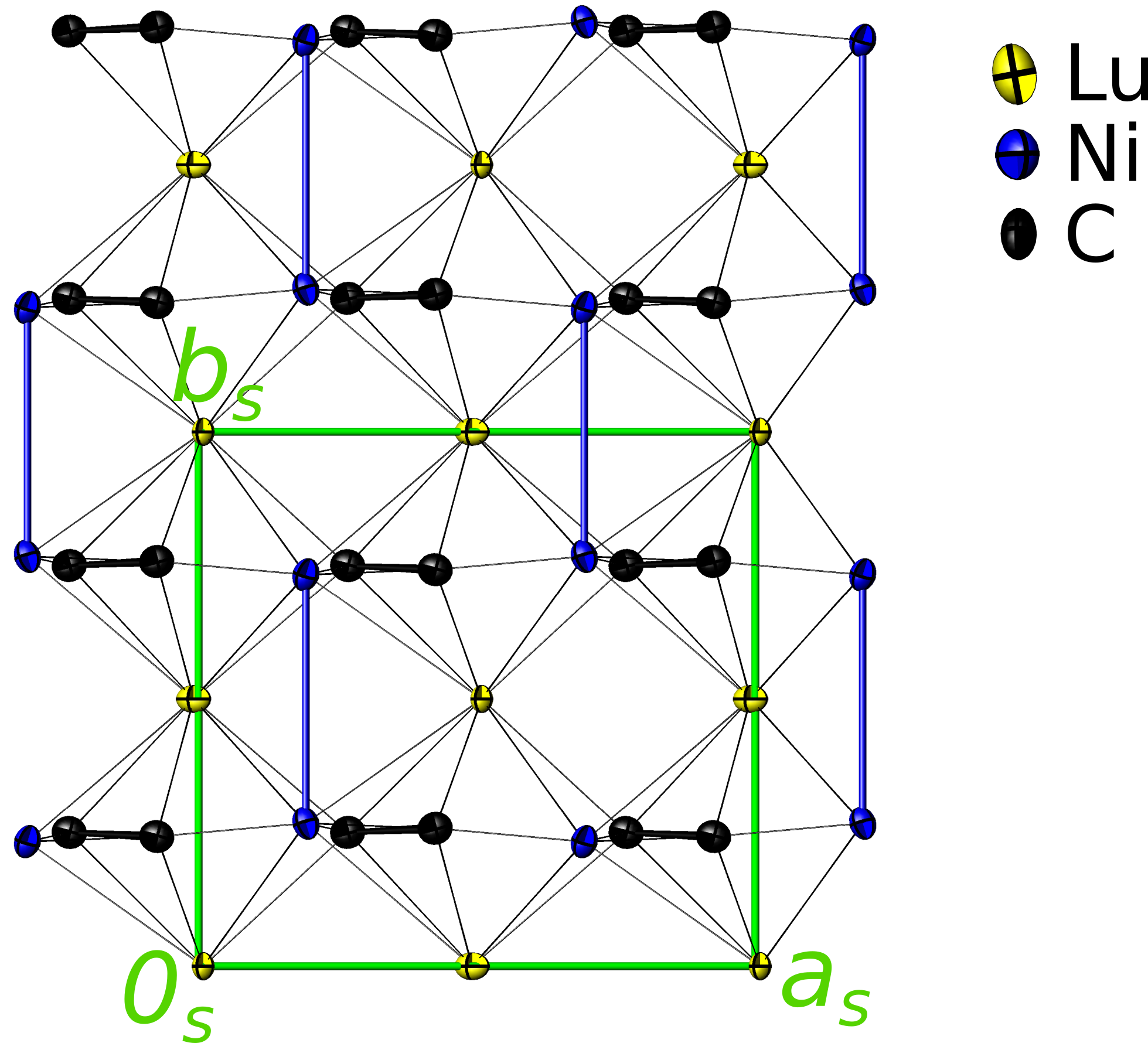}
			\caption{Projection of the CDW superstructure as view along the monoclinicic $b$-axis 
			($||a$-axis of the orthorhombic cell) in the upper panel
			indicates the orthorhombic and monoclinic settings as red and green cells, respectively. 
			The view along the monoclinicic $c$-axis in the lower panel highlights the CDW superstructure modulation
			of the Ni-Ni distances.}
	\label{CDWmodulation}
	\end{center}
\end{figure}

To structurally characterize the CDW order, the superstructure reflections of a LuNiC$_2$ data set collected at 100\,K 
and up to high diffraction angles ($2\theta = 100^\circ$) were analyzed. Indexed in the reciprocal basis of the $Amm2$
parent structure, these superstructure reflections are located at 
$h=n_h+\frac1 2$, $k=n_k+\frac1 2$, $l=n_l+\frac1 2$, $n_h,n_k,n_l\in\mathbb Z$
(Fig.~\ref{reci}). 
They can be attributed to two monoclinic C-centered ($mC$) domains, with the standard centered basis
$(\mathbf b+\mathbf c,2\mathbf a,(\mathbf b-\mathbf c)/2)$ with respect to the parent structure. The monoclinic
direction of the $mC$ domains corresponds to the $a$-axis of the orthorhombic parent structure. Assuming a
symmetry reduction of the group/subgroup kind, only the $Cm$ and $Cc$ space groups are possible.
$Cc$ is ruled out owing to strong violations of the corresponding systematic absences. Indeed, only in the
$Cm$ space group reasonable refinements were obtained.

The lost orthorhombic symmetry is retained as twin law (twin operations $m_{[010]}$, $2_{[001]}$ with
respect to the orthorhombic basis). Whereas the main reflections of both domains overlap, the superstructure
reflections form two disjoint sets (twinning by reticular pseudo-merohedry, Fig.~\ref{reci}).
A model of the CDW superstructure was constructed by symmetry reduction from the $Amm2$ model
and refined against ``HKLF5'' data with information on reflection overlap.
Crystal and refinement data as well as final coordinates are listed in Table~\ref{tab2}.

The symmetry reduction from the $Amm2$ parent- to the $Cm$ superstructure is of index four and can be decomposed
into two steps, of the \emph{translationengleiche} (reduction of point symmetry) and the \emph{klassengleiche}
(reduction of translation symmetry) type, respectively.
The Lu position in the CeNiC$_2$-type parent structure with site symmetry $mm2$ is split in two positions with $m$ site symmetry.
The C position ($m..$) is split in two general positions.
The site symmetry of the Ni atom is reduced from $mm2$ to $1$, without being split.
Consistent with the refinement of the parent structure (see above, Table~\ref{tab1}), 
the largest deviation from the $Amm2$ symmetry is observed for the Ni atoms.
They are displaced distinctly along the monoclinic $b_{(s)}$-axis (being parallel to the orthorhombic $a$-axis of the parent structure),
thus, forming pairs of Ni atoms related by the mirror reflection with a shortened Ni-Ni distance of 3.208(2) \AA{} 
as compared to $b_{(s)}/2 = 3.445(1)$ \AA. 
The large and highly anisotropic displacement parameters of Ni in the $Amm2$ averaged structure model 
become smaller and more isotropic in the $Cm$ superstructure, 
i.e.\ comparing the anisotropic displacement matrix after main axis transformation, 
$U_{ii}$/\AA$^2$ change from (0.0217,0.0068,0.0060) to (0.0077,0.0057,0.0042), respectively. 
The equidistant periodicity of the Ni atoms in the orthorhombic parent structure is, thus, 
replaced by alternating shortened and elongated Ni-Ni distances (3.208 vs.\ 3.682~\AA{})
which resemble a Peierls-type distortion, also referred to as a Peierls dimerization.  
The corresponding model of the CDW modulation of LuNiC$_2$ is depicted in Fig.~\ref{CDWmodulation} as 
views along the monoclinic $b_{(s)}$-axis in the upper and along the monoclinic $c_{(s)}$-axis in the lower panel.
The latter emphasizes the Peierls-type distortion of the Ni atom periodicity along the $b_{(s)}$-axis, 
via fat blue solid lines highlighting the pairwise shortened Ni-Ni distances of 3.208(2)~\AA.
The CDW modulation of the orthorhombic parent structure corresponds to a modulation wave vector 
$\mathbf{Q}=(\frac1 2 , 0 , \frac1 2)$ in the IRBZ as displayed in Fig.~\ref{BZ}.

\begin{figure}[t]
	\begin{center}
		\includegraphics[width=0.95\columnwidth]{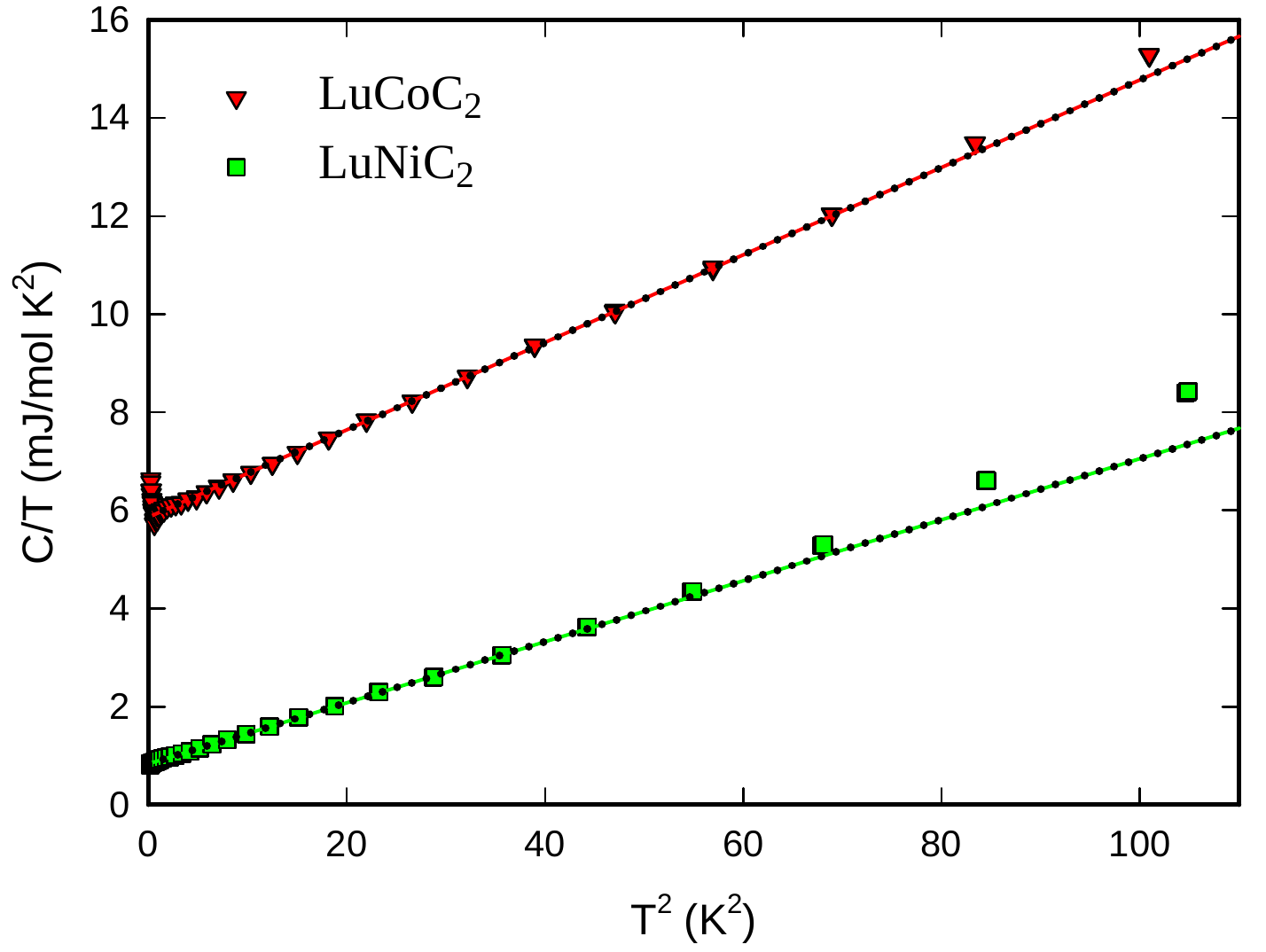}
		\caption{Low temperature specific heat as $C/T$ vs. $T^2$ of single crystals LuCoC$_2$ and LuNiC$_2$ as labeled; 
		dotted lines are linear fits for limited temperature intervals (see text).}
	\label{fi_cp}
	\end{center}
\end{figure}

\subsection{Specific heat, magnetic susceptibility and electrical resistivity of LuCoC$_2$ and LuNiC$_2$}
\label{Cp_res}

The specific heat data displayed in Fig.~\ref{fi_cp} reveal for both, LuCoC$_2$ and LuNiC$_2$, 
a metallic behavior, i.e.\ $C(T)\simeq\gamma T+(12\pi^4n/5)(T/\Theta_{\rm D})^3$ at low temperatures,
where $\gamma$ represents the Sommerfeld coefficients of the $T$-linear electronic specific heat contribution, 
$n$ is the number of atoms in the formula unit (i.e.\ $n=4$ for LuCoC$_2$ and LuNiC$_2$), 
and $\Theta_{\rm D}$ is the Debye temperature, characterizing the low temperature lattice
heat capacity. LuCoC$_2$ displays a distinct up-turn in $C/T$ at lowest temperatures (400\,--\,600\,mK), 
which is similar to the one observed for elemental lutetium \cite{PhysRev.133.A219}. 
Such low temperature up-turn in $C/T$ remains, at least for $T>0.4$~K, absent for LuNiC$_2$.     
The corresponding linear fits of the data (see the dotted lines in Fig.~\ref{fi_cp}) are applied
to temperature intervals, $3< T^2<60$\,K$^2$ for LuCoC$_2$ and $T^2<40$\,K$^2$ for LuNiC$_2$, thus, 
yielding the Sommerfeld coefficients, $\gamma=5.9$(1)\,mJ/mol\,K$^2$ and 0.83(5)\,mJ/mol\,K$^2$ 
as well as Debye temperatures, $\Theta_{\rm D}=444$\,(8)\,K and 500\,(10)\,K, respectively.
While phonon specific heat contributions of LuCoC$_2$ and LuNiC$_2$ are closely matching 
each other in an extended temperature interval (not shown), electronic contributions of 
LuCoC$_2$ and LuNiC$_2$ and, thus, their electronic density of states at the Fermi level, $N(E_f)$, 
are strikingly different.
For both, LuCoC$_2$ and LuNiC$_2$, an anomaly, which could be indicative of bulk superconductivity, remains absent.

Field dependent magnetization and temperature dependent magnetic susceptibility measurements (not shown) reveal even for 
single crystalline LuNiC$_2$, prepared by the floating zone technique, traces of a ferromagnetic 
impurity phase with $T_{\rm C}\sim 80$\,K ($\mu_{\rm sat}\sim 2\times 10^{-10}\mu_{\rm B}/$f.u.) 
and of a superconducting impurity phase with $T_{c}\sim 5$\,K. 
The isothermal magnetization, $M(H)$ measured at 2 K, is characteristic of a hard type-II superconductor 
with $H_{c1}\sim 5$\,mT and $H_{c2}\sim 1$\,T. The diamagnetic Meissner volume fraction is only about 
$5\times 10^{-3}$ and, thus, attributed to a superconducting impurity phase precipitated in the 
zone refined crystal (compare SEM results in section~\ref{ED}). 
The intrinsic magnetic susceptibility of LuNiC$_2$ is diamagnetic with
$\chi\simeq -2.2\times 10^{-5}$ cm$^3$/mol at room temperature, which is roughly two third of the expected 
core-diamagnetic susceptibility  $\chi_{\rm core}\sim -3\times 10^{-5}$ cm$^3$/mol (see Ref.~\cite{selwood}).
Both, the ferromagnetic and the superconducting impurity phases, are obviously well dispersed in the single 
crystalline matrix in form of small inclusions (see above).

\begin{figure}[t]
	\begin{center}
		\includegraphics[width=0.95\columnwidth]{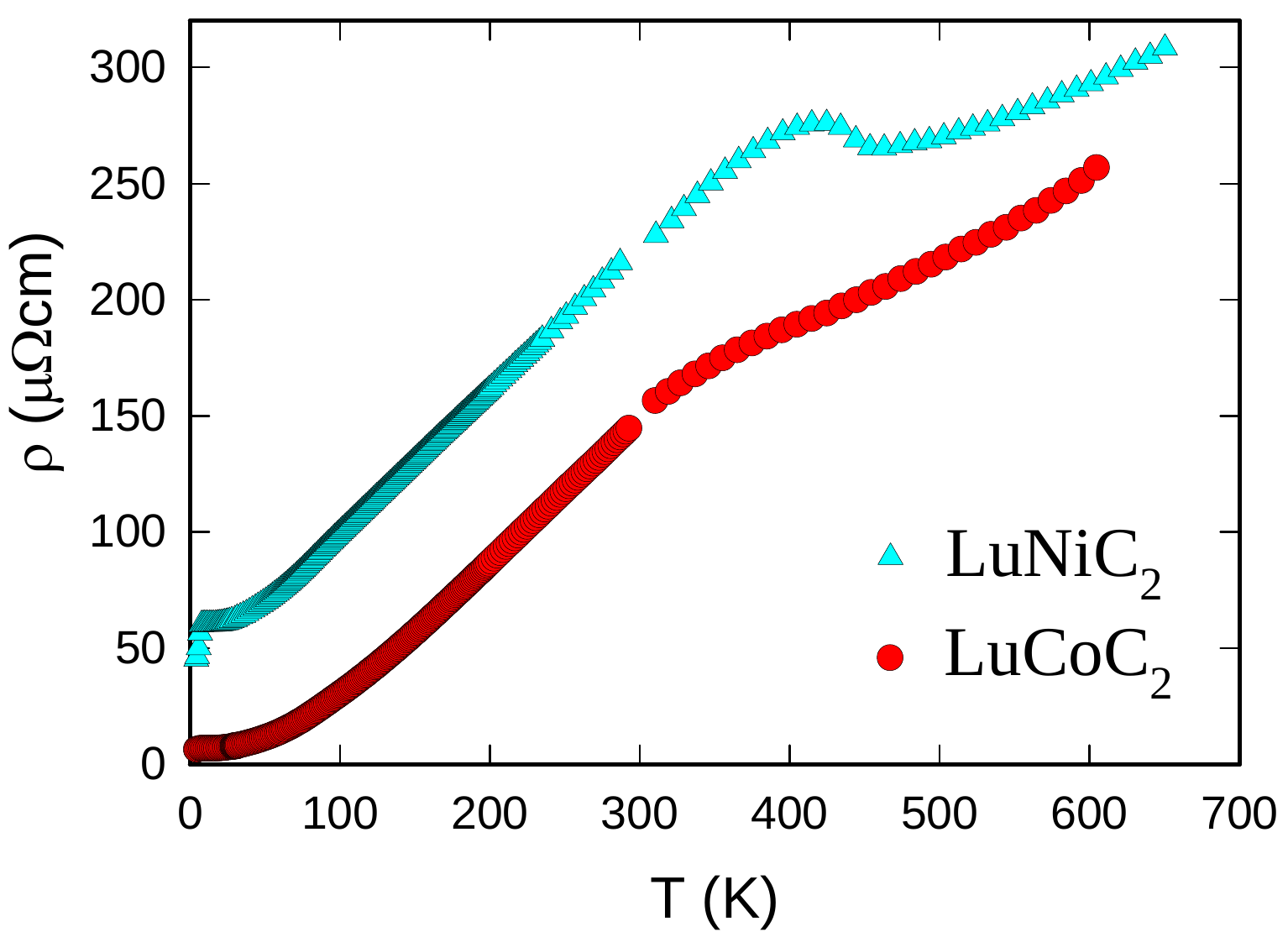}
		\caption{Temperature dependent electrical resistivity, $\rho(T),$ of 
		polycrystalline samples LuCoC$_2$ and LuNiC$_2$ as labeled.}
	\label{fi_res}
	\end{center}
\end{figure}

\begin{figure}[t]
	\begin{center}
		\includegraphics[width=0.95\columnwidth]{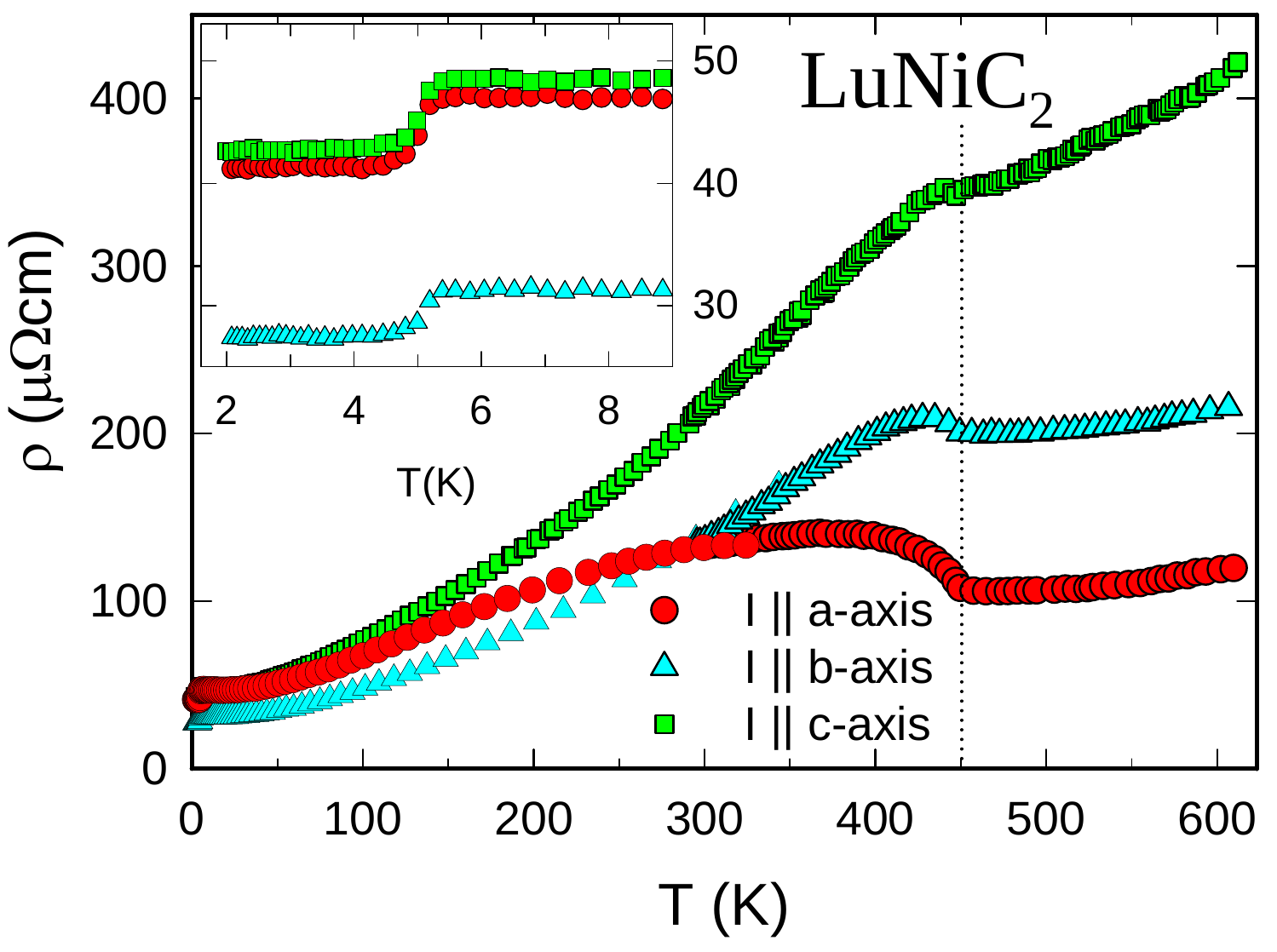}
		\caption{Temperature dependent electrical resistivity, $\rho(T),$ of LuNiC$_2$ single crystals
		with the electric current applied parallel to crystal orientations as labeled. 
		The dotted line indicates the onset of CDW order at 447\,(5)~K.}
	\label{fi_SC}
	\end{center}
\end{figure}

Temperature dependent electrical resistivity measurements on polycrystalline samples displayed in Fig.~\ref{fi_res} 
reveal a simple metallic behavior of LuCoC$_2$ with a reasonably low residual resistivity, $\rho_0\simeq 7$\,$\mu\Omega$cm, 
whereas for LuNiC$_2$ the residual resistivity, $\rho_0\sim 50$\,$\mu\Omega$cm, is largely enhanced and two distinct 
anomalies are observed at 7\,K and near 450\,K. 
The latter resembles the resistive CDW anomalies of other $R$NiC$_2$ compounds (with $R=$ Pr\,--\,Tb), 
initially reported by Murase {\sl et al.}~\cite{murase}, and is well in line with the trend of CDW transition 
temperatures, $T_{\rm CDW}$,  discussed in Ref.~\cite{PhysRevB.97.041103} where $T_{\rm CDW}=463$\,K is suggested from
preliminary results of LuNiC$_2$ which are in close agreement with the present data.  
The resistive drop observed at 7\,K does not relate to any traceable, correspondent anomaly in the heat capacity, but
does relate to a spurious diamagnetic signal of the magnetic susceptibility (see above) and is, thus, attributed to a 
small superconducting impurity fraction which is well below the percolation limit.

The anisotropy of the electrical resistivity of single crystalline LuNiC$_2$, 
displayed in Fig.~\ref{fi_SC}, is obtained from measurements with current applied 
along principal crystallographic orientations. 
The dotted vertical line in Fig.~\ref{fi_SC} marks the onset of CDW order
at $T_{\rm CDW}=447$\,(5)\,K. 
At temperatures exceeding the CDW transition, i.e.\ $T>T_{\rm CDW}$, 
LuNiC$_2$ exhibits a similar anisotropy
of the resistivity, $\rho_a<\rho_b<\rho_c$, as e.g.\  reported earlier for 
SmNiC$_2$, GdNiC$_2$, and TbNiC$_2$~\cite{shimomura}. 
At lower temperatures ($T<T_{\rm CDW}$), however, LuNiC$_2$ behaves distinctly different
than magnetic $R$NiC$_2$ single crystals investigated by Shimomura 
{\sl et al.}~\cite{shimomura}:
(i) the CDW anomaly below $T_{\rm CDW}$ is most pronounced for $\rho_a$,
while those of hitherto investigated magnetic $R$NiC$_2$ is most prominent for $\rho_c$ and 
(ii) LuNiC$_2$ displays an almost isotropic electrical resistivity at $T<50$\,K ($\rho_a\sim 1.5\rho_b\sim\rho_c$), 
whereas $\rho_a\ll\rho_b<\rho_c$ has been reported for magnetic $R$NiC$_2$.
Finally (iii), low temperature resistive anomalies of the latter relate to the onset of
rare earth magnetism and the resulting modifications (suppression) of CDW order.
LuNiC$_2$ displays a resistive anomaly near 5\,K which is suppressed by an 
externally applied magnetic field (not shown). The critical field for suppressing the 
roughly isotropic resistivity drop by a few $\mu\Omega$cm
is (independent of orientation) about 1\,T at 2\,K and decreases as a function
of temperature as typical for a superconducting phase (see the above discussion of 
magnetization data).

\begin{figure}[t] 
\begin{center}
\includegraphics[width=0.98\columnwidth]{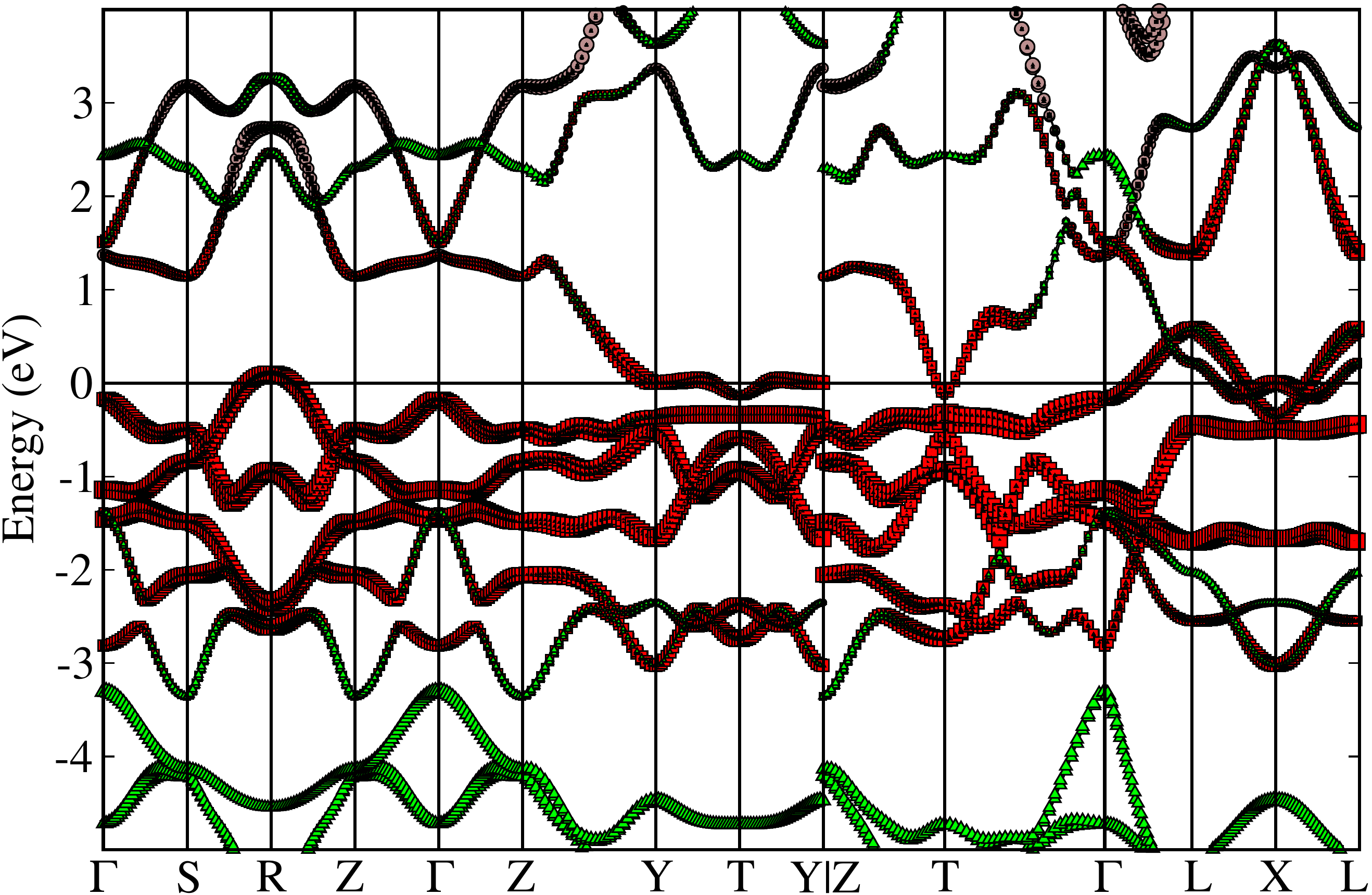}
\includegraphics[width=0.98\columnwidth]{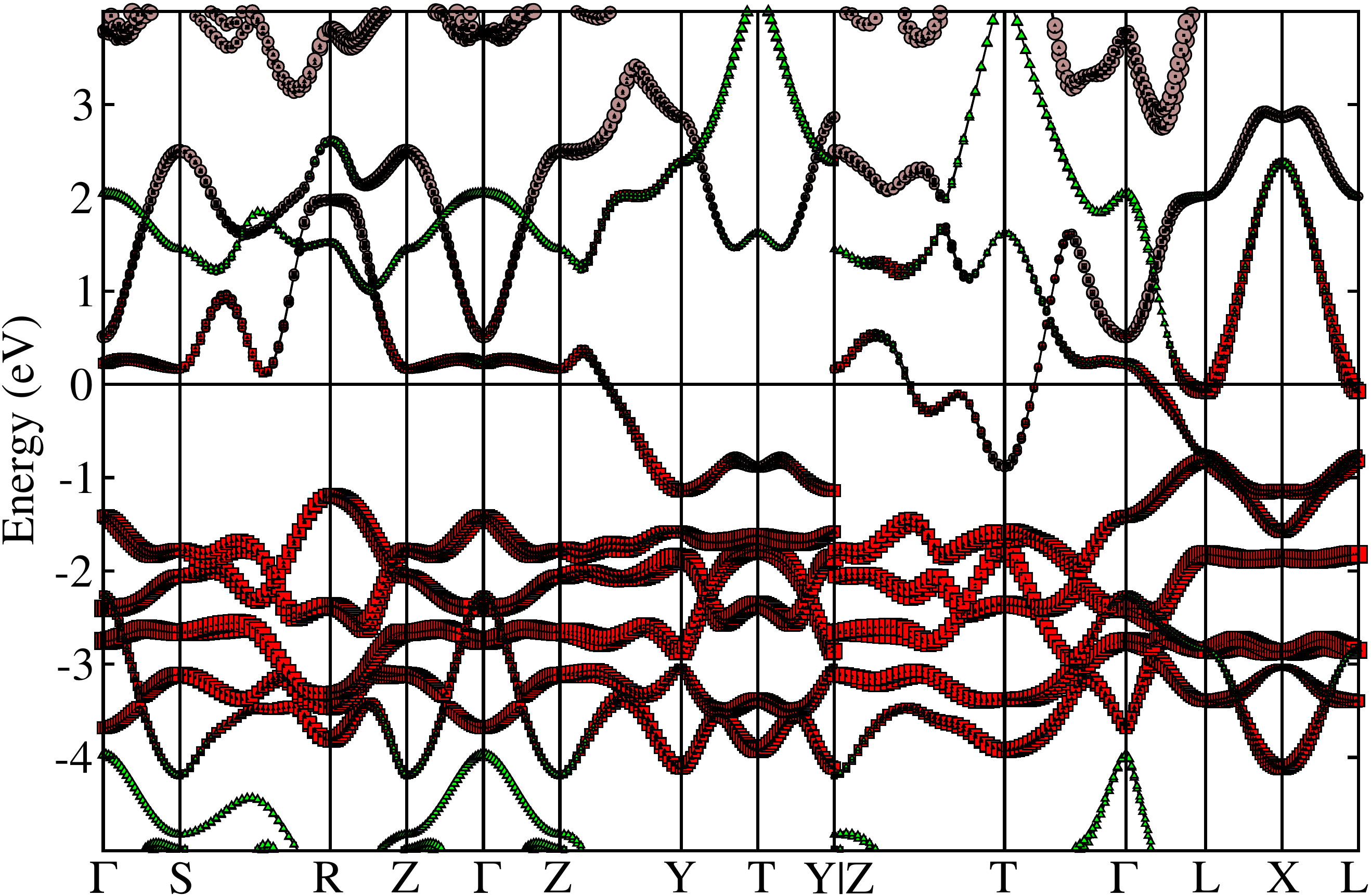}
\caption{Calculated electronic band structure of LuCoC$_2$ (upper panel) and LuNiC$_2$ (lower panel).
In both plots brown circles denote a Lu character of the bands, green triangles denote a C related origin.
In the upper panel red squares refer to Co $3d$ dominated bands and in the lower panel 
to Ni $3d$ dominated bands. 
Zero energy (indicated by the horizontal line) refers to the position of the Fermi level.
The diameter of the symbols shows the spectral weight related to each eigenstate 
of the system at a given $k$-point.} 
\label{bs}
\end{center}
\end{figure}

\begin{figure}[t] 
\begin{center}
\includegraphics[width=0.92\columnwidth]{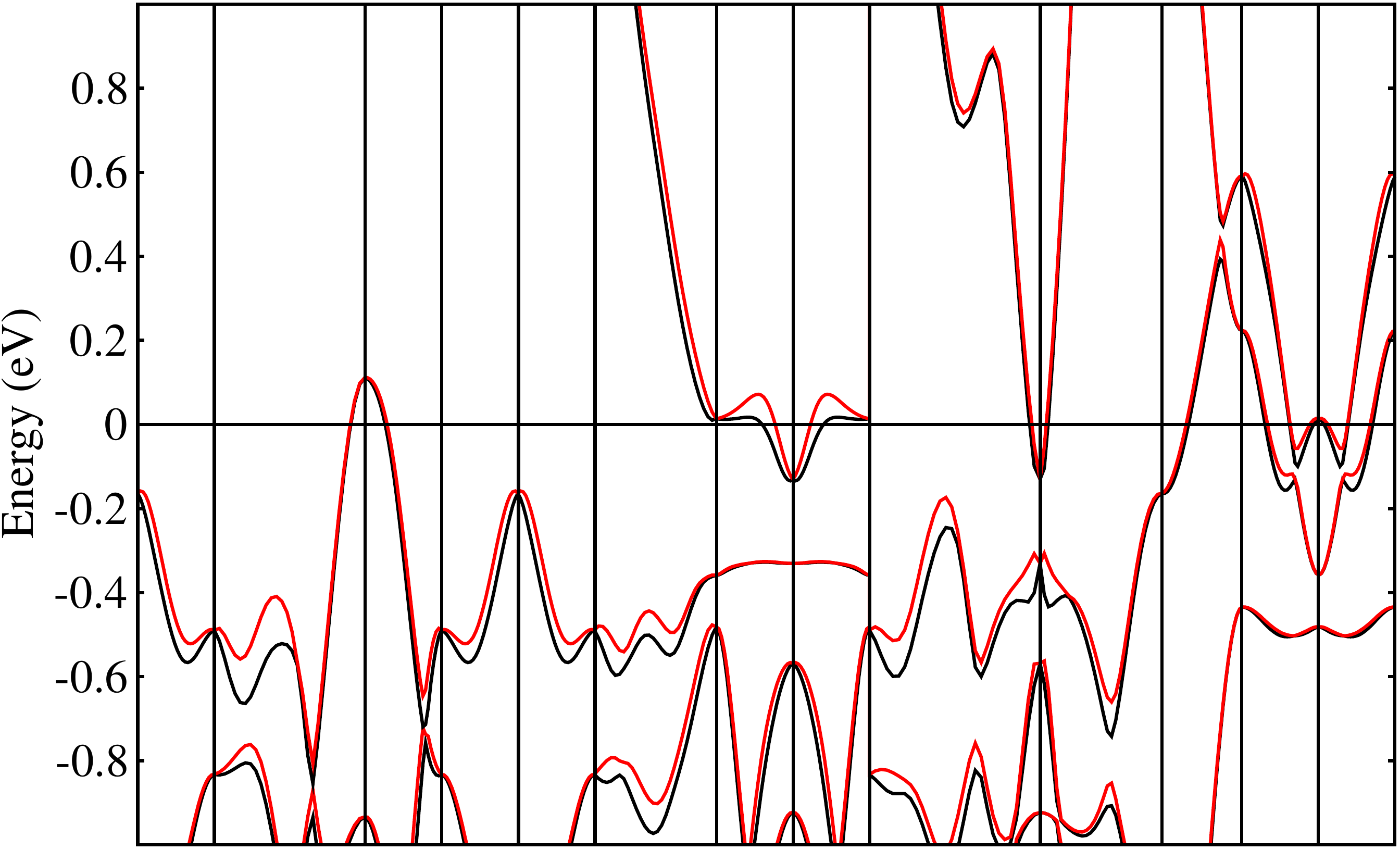}
\includegraphics[width=0.92\columnwidth]{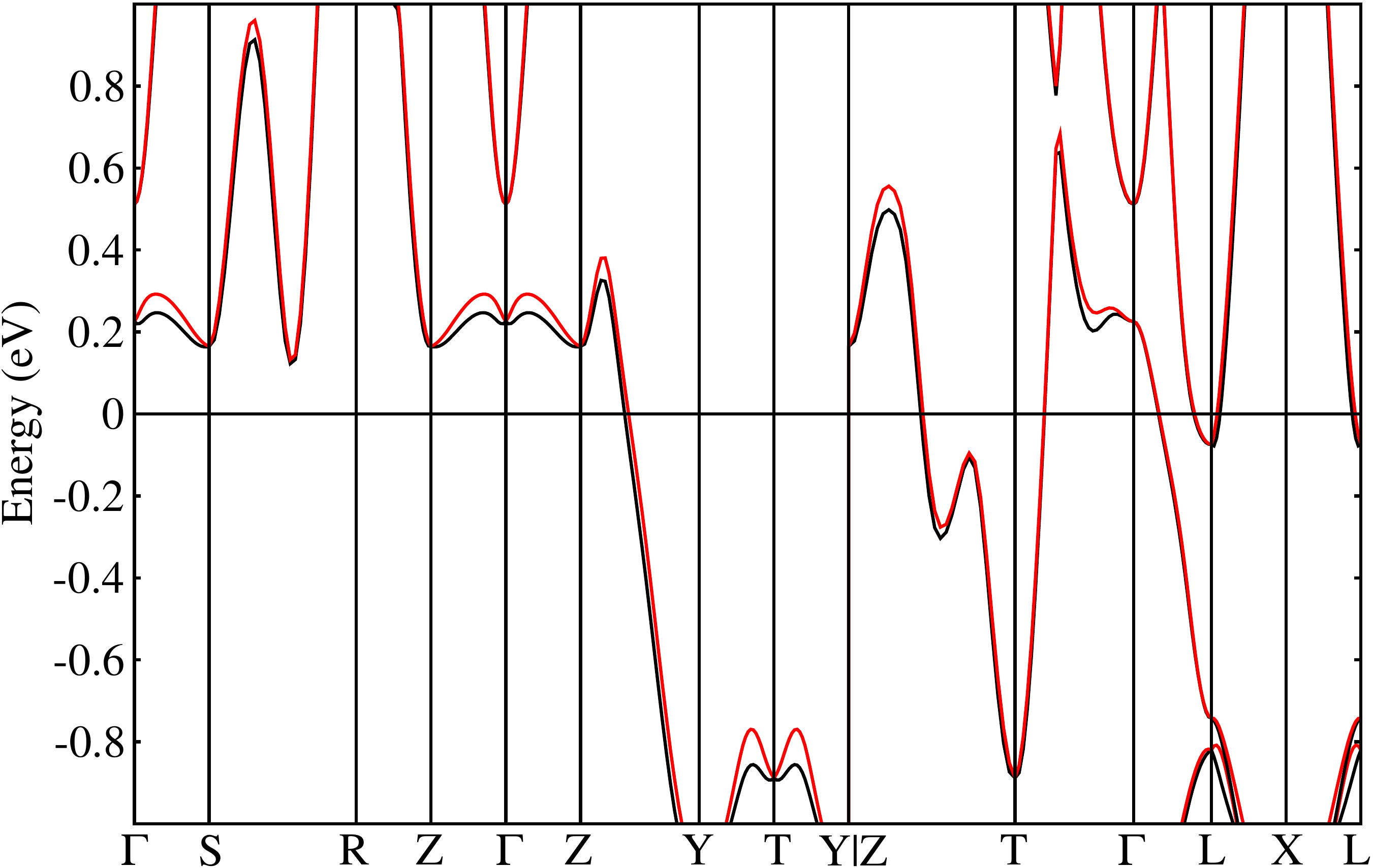}
\caption{Calculated electronic band structure of LuCoC$_2$ (upper panel) and LuNiC$_2$ (lower panel).
Here a zoom of the BS is shown. The red and black curves refer to the two spin channels of SOC.} 
\label{bs2}
\end{center}
\end{figure}

\subsection{Electronic structure studies of LuCoC$_2$ and LuNiC$_2$}
\label{ESS}

\subsubsection{Results based on the CeNiC$_2$-type orthorhombic structure model}
\label{ESSa}

Figure~\ref{bs} displays the calculated electronic band structure of LuCoC$_2$ and LuNiC$_2$ for high-symmetry directions in the 
first Brillouin zone (compare Fig.~\ref{BZ}). 
The dispersions are shown as a fat-band plot, which is a band structure equivalent to a projected EDOS plot.
Thereby, symbols (brown circles for Lu site projected states, green triangles for C states, and red squares for the $d$-metal Co- and Ni-states) 
refer to the dominant atomic origin of the crystal orbitals and the widths of all bands are proportional to the relative contributions of a 
given set of atomic orbitals to the crystal orbitals. 
The upper panel of Fig.~\ref{bs} reveals for LuCoC$_2$ that two Co $3d$ bands are crossing the Fermi energy.
There is only some dispersion of these bands at $\Gamma$ and $R$, else these bands are flat. 
Above the Fermi level, bands origin from mixed contributions from Co, C and Lu 
and exhibit, in part, strong dispersion.

For LuNiC$_2$  (lower panel of Fig.~\ref{bs}) a shift of the Fermi level by about 1\,eV is observed
(with respect to LuCoC$_2$), which 
almost resembles the expected filling of electronic states in terms of a rigid band picture.
In the case of LuNiC$_2$, essentially one band with a mixed Lu-Ni-C character crosses the Fermi energy while the 
almost filled $3d$ dominated bands cross the Fermi level only near the $L$ point located at the IRBZ boundary (compare
Fig.~\ref{BZ}) and originate a small electron pocket of the Fermi surface around this place 
(see below for further discussions of the Fermi surface).

In order to uncover the band splitting due to the asymmetric SOC caused by the 
lack of inversion symmetry of the crystal structure, Fig.~\ref{bs2} displays a closer view of the band structure 
in the vicinity of the Fermi level, where red and black lines refer to the two spin channels of SOC
in the absence of inversion symmetry of the crystal structure. 
The largest splittings are observed for $3d$ dominated bands and, with respect to band crossing the 
Fermi level, spin-orbit splitting is most pronounced for LuCoC$_2$ while for LuNiC$_2$, smaller band splittings 
due to SOC are observed.

\begin{figure}[t] 
\begin{center}
\includegraphics[width=0.98\columnwidth]{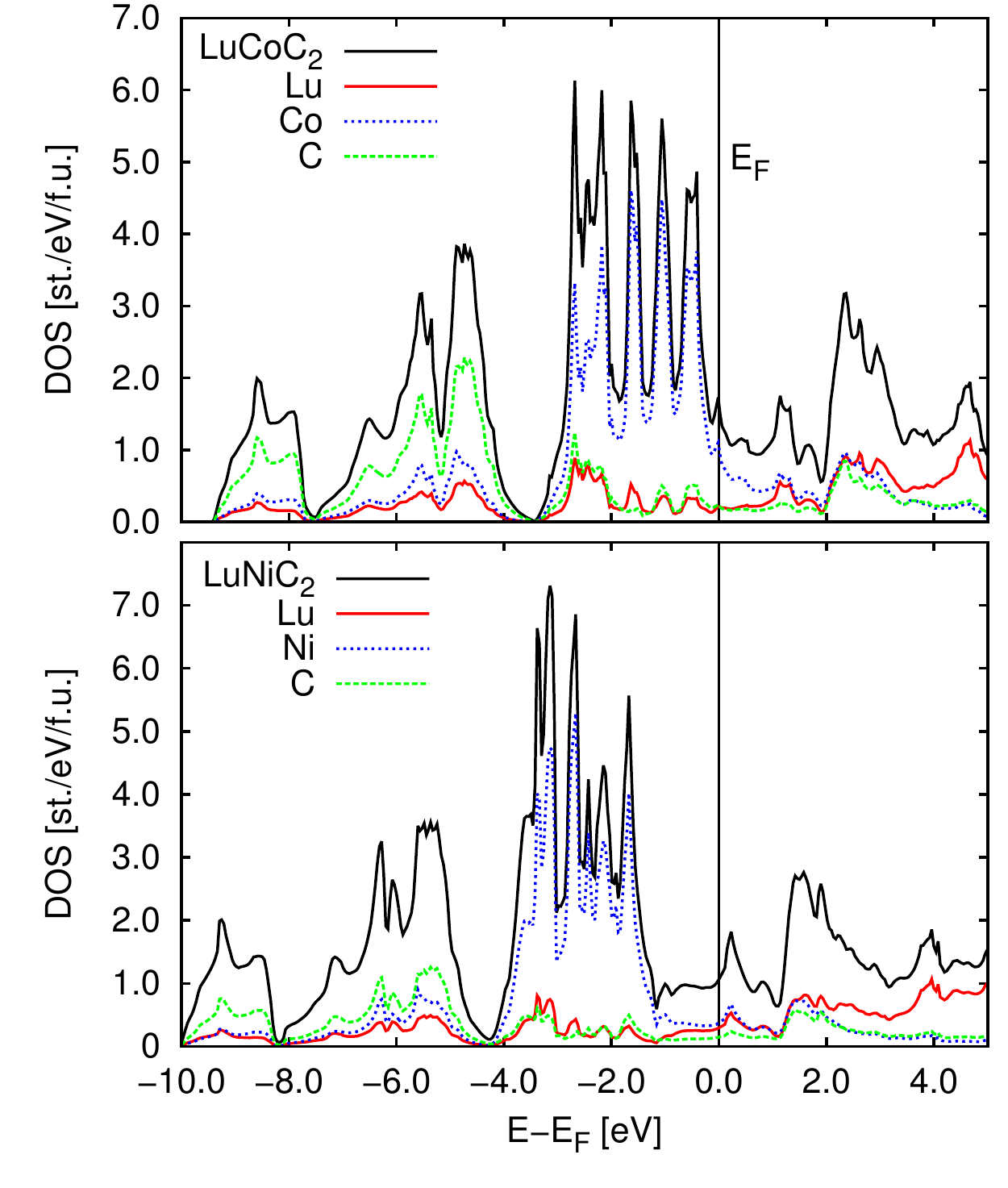}
\caption{Calculated electronic density of states for LuCoC$_2$ (upper panel) and LuNiC$_2$ (lower panel).
The total EDOS is shown as a black solid line and the atom projected EDOS for Lu, Co and C sites are 
indicated as red solid, blue dotted, and green dashed lines, respectively.}
\label{dos}
\end{center}
\end{figure}

Figure~\ref{dos} presents the calculated EDOS of LuCoC$_2$ and LuNiC$_2$ in the upper and lower panel, respectively.
Both EDOS plots have very similar features, except for the position of the Fermi energy, which relates 
to the different $3d$ electron count of Co as compared to Ni. The EDOS at the Fermi energy, 
$N(E_{\rm F})$, of LuCoC$_2$ and LuNiC$_2$ are $1.62$ states/eV\,f.u.\ and $1.03$ states/eV\,f.u., 
respectively. The corresponding site projected EDOS contributions at the Fermi level are dominated by about 80\%
by Co $3d$ contributions of LuCoC$_2$, whereas for LuNiC$_2$, there are almost equal contributions to $N(E_{\rm F})$ 
from Ni and Lu and still relevant ones from carbon projected states. 

\begin{figure}[t] 
\begin{center}
\includegraphics[width=0.85\columnwidth]{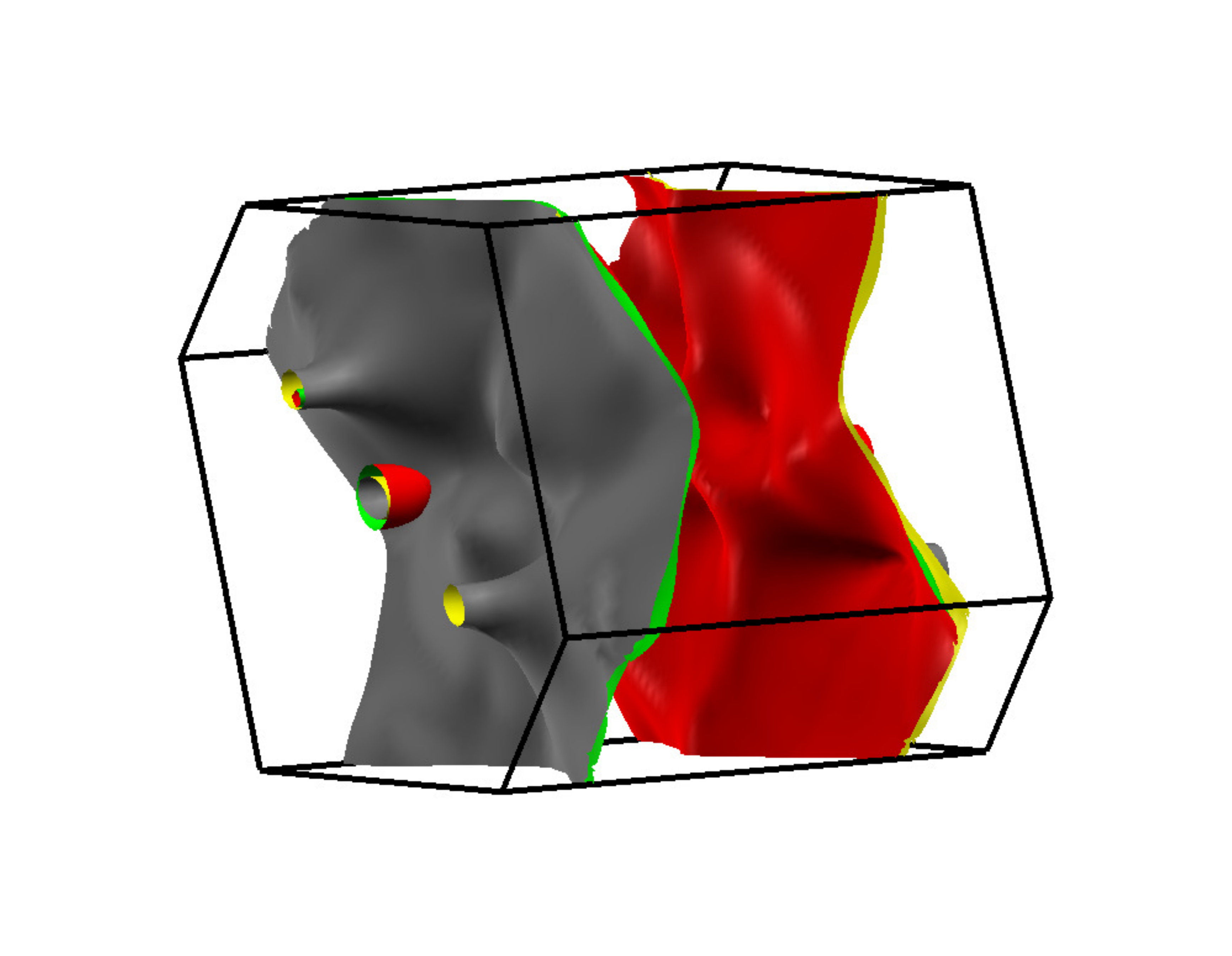}
\caption{The Fermi surfaces of LuNiC$_2$ displayed with orientation $b_1$ horizontally 
and $b_2$ vertically, as indicated by the first Brillouin zone boundaries.
The two spin channels of SOC are indicated by colored sheets where red/yellow sides are towards unoccupied and green/gray 
sides are towards occupied states.}
\label{fs}
\end{center}
\end{figure}

The Fermi surface of LuNiC$_2$,  depicted in Fig.~\ref{fs}, consists of 
two sheets which are each moderately split by SOC into sheets related to the
two spin channel of SOC. 
One is a pair of large quasi-planar sheets, which are connected by two neckings 
at the IRBZ boundary along the $b_1$ direction and the second is a small electron pocket centered
at the $L$-point of the IRBZ (compare Fig.~\ref{BZ}).
\begin{figure}[t] 
\begin{center}
\includegraphics[width=0.49\columnwidth]{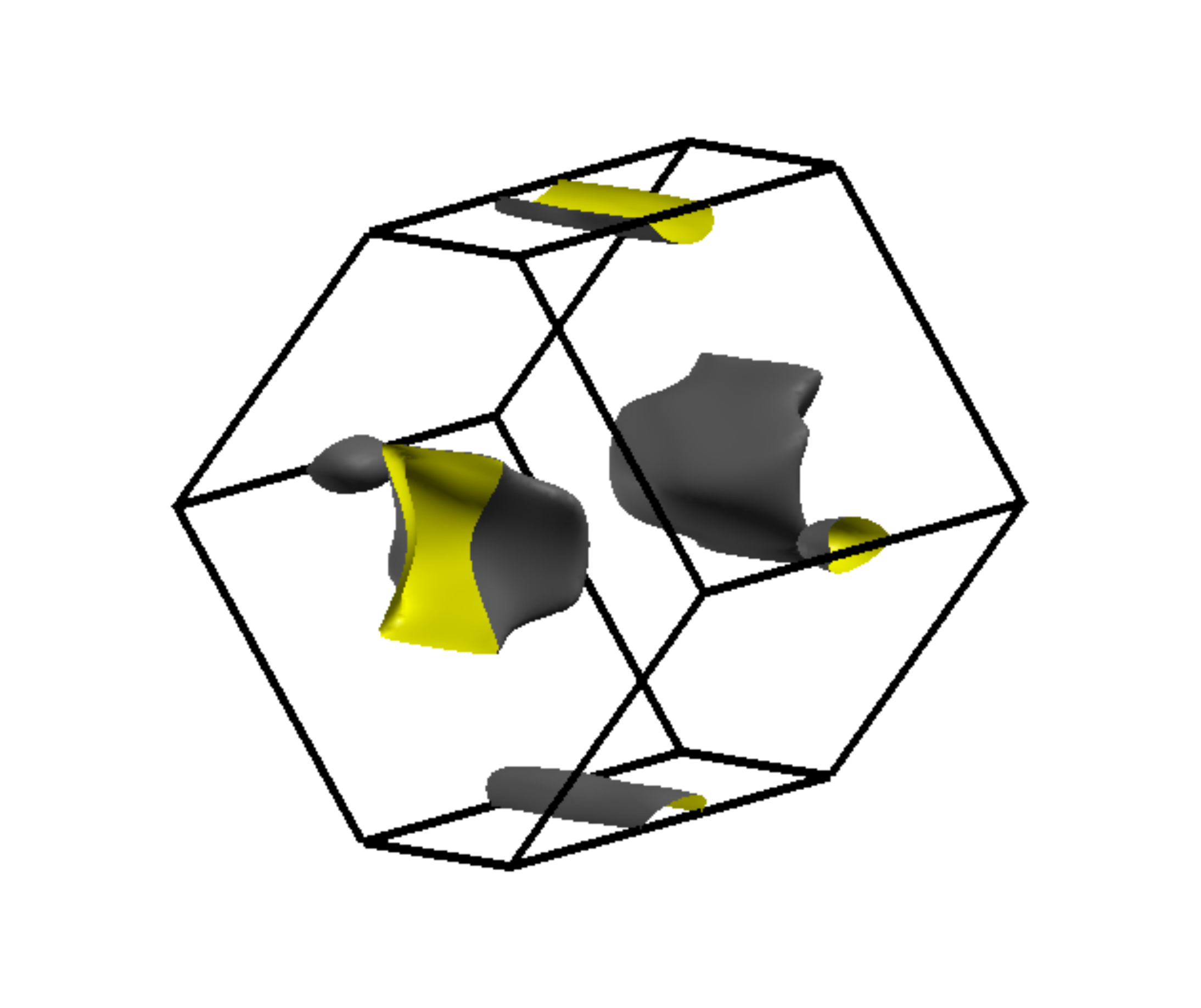}
\includegraphics[width=0.49\columnwidth]{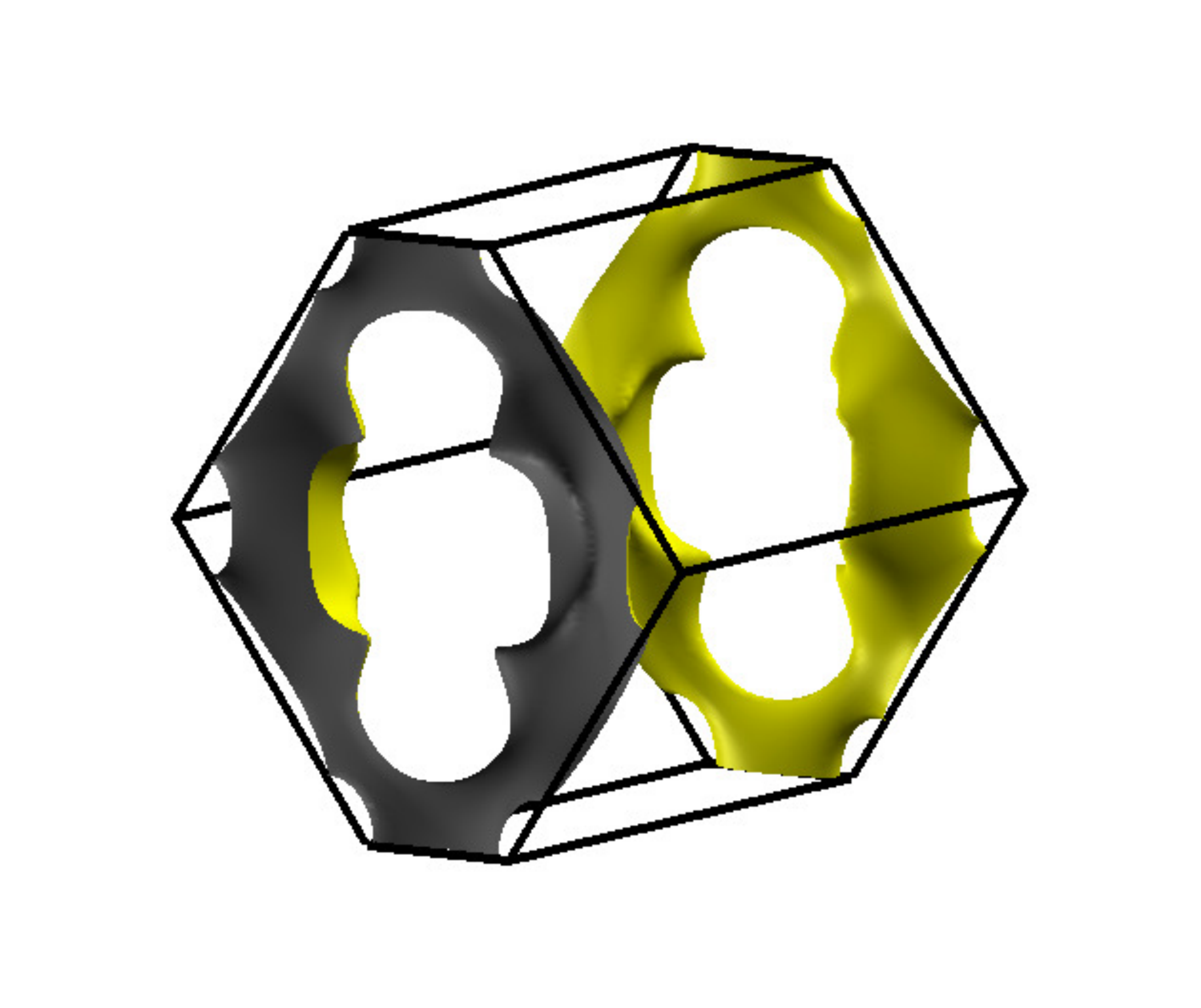}
\caption{Fermi surfaces of LuCoC$_2$ are depicted separately for two bands
(neglecting SOC, see text): 
in the left panel, band~$16$ is displayed, and in the right panel, band~$17$. 
Gray and yellow sides face to occupied and unoccupied states, respectively.}
\label{fs2}
\end{center}
\end{figure}
The larger Fermi surface sheet approximates to that of a quasi-one-dimensional electronic band
and, with evolving modifications within the series LaNiC$_2$ to LuNiC$_2$, seems typical for $R$NiC$_2$ compounds.
Calculations for early rare earth members, LaNiC$_2$ and SmNiC$_2$, by Laverock {\sl et al.}~\cite{laverock}
suggested a similar pair of large sheets oriented perpendicular to the $b_1$ direction which, however,
display an additional necking at the position of the electron pocket of LuNiC$_2$.
On the contrary to LuNiC$_2$, LaNiC$_2$ is reported in Ref.~\cite{laverock} to display a significant
electron pocket centered at the $\Gamma$-point, which becomes small for SmNiC$_2$ and is absent in the 
present calculation of LuNiC$_2$ and also in the Fermi surface calculation of YNiC$_2$ reported by Hase and Yanagisawa~\cite{IzumiHase09}. 
The latter results are in rather close match with the present ones of LuNiC$_2$ presented in Fig.~\ref{fs}.

\begin{figure}[t] 
\begin{center}
\includegraphics[width=1.0\columnwidth]{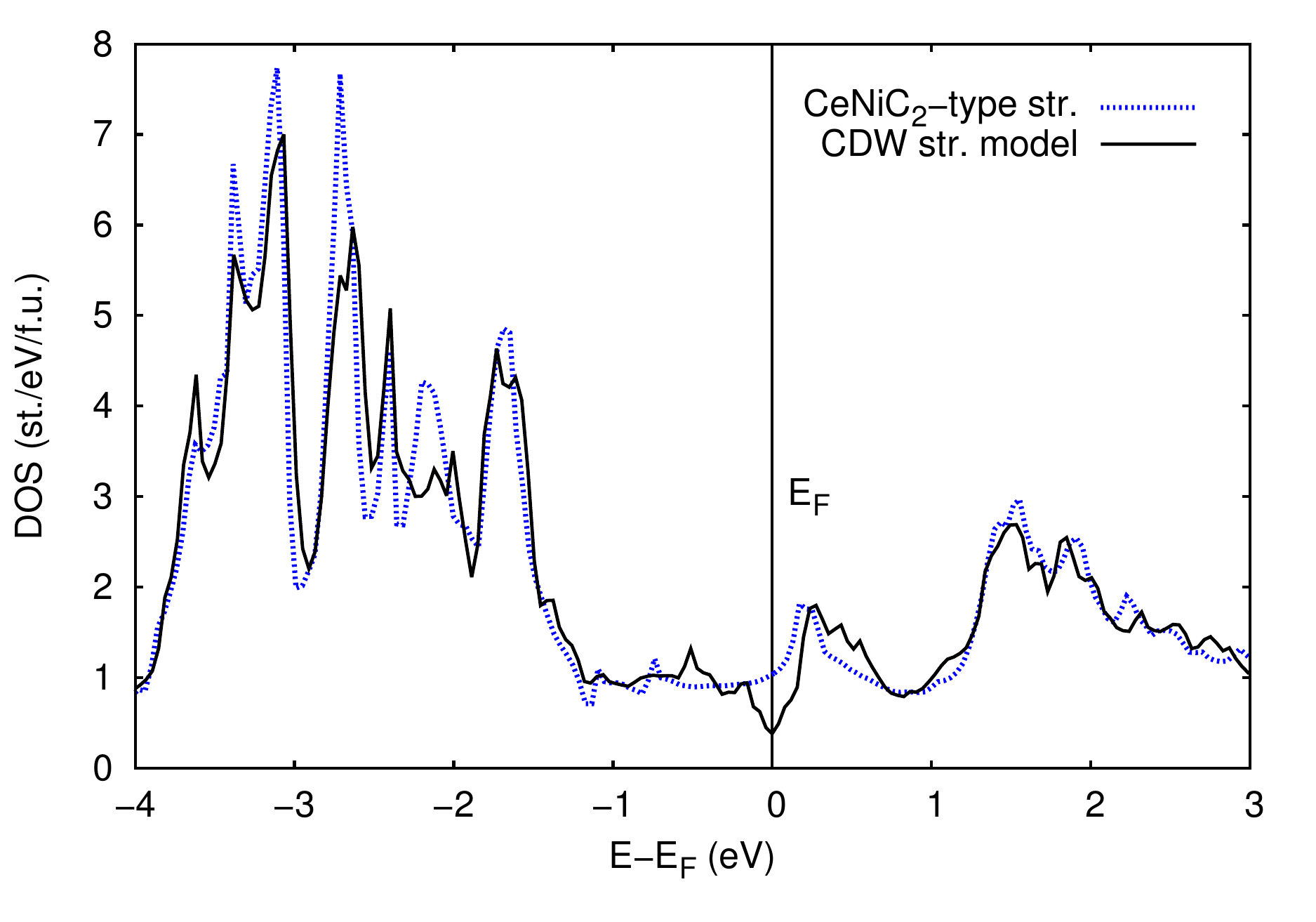}
\caption{Calculated electronic density of states of LuNiC$_2$ based on its structure models with 
and without CDW superstructure is shown as black solid line and blue dotted line, respectively.}
\label{dos_cdw}
\end{center}
\end{figure}

Calculations of LuCoC$_2$ reveal the two Fermi surface sheets depicted in Fig.~\ref{fs2}, where that 
of band 17 in the right panel corresponds to the large quasi-planar Fermi surface sheets of LuNiC$_2$, 
however, at a much lower band-filling state resulting from the lower $d$-electron 
count of Co as compared to Ni. For the sake of clarity and focus on the essential features of the Fermi surface, 
the spin splitting due the asymmetric SOC is neglected in Fig.~\ref{fs2}.

\subsubsection{Comparison of LuNiC$_2$ results based on structure models with and without CDW modulation}
\label{ESSb}

Figure~\ref{dos_cdw} presents a direct comparison of the EDOS calculated for two structure models of 
LuNiC$_2$, the CeNiC$_{2}$-type parent structure and the CDW modulated structure, which
reveals the formation of a partial CDW gap right at the Fermi energy.
The value of the EDOS at the Fermi energy, $N(E_{\rm F})=0.38$ states/eV\,f.u., of the CDW modulated structure is,
thus, significantly reduced as compared to $N(E_{\rm F})=1.03$ states/eV\,f.u.\ obtained for the orthorhombic 
parent structure of LuNiC$_2$. 

\begin{figure}[t] 
\begin{center}
\includegraphics[width=0.89\columnwidth]{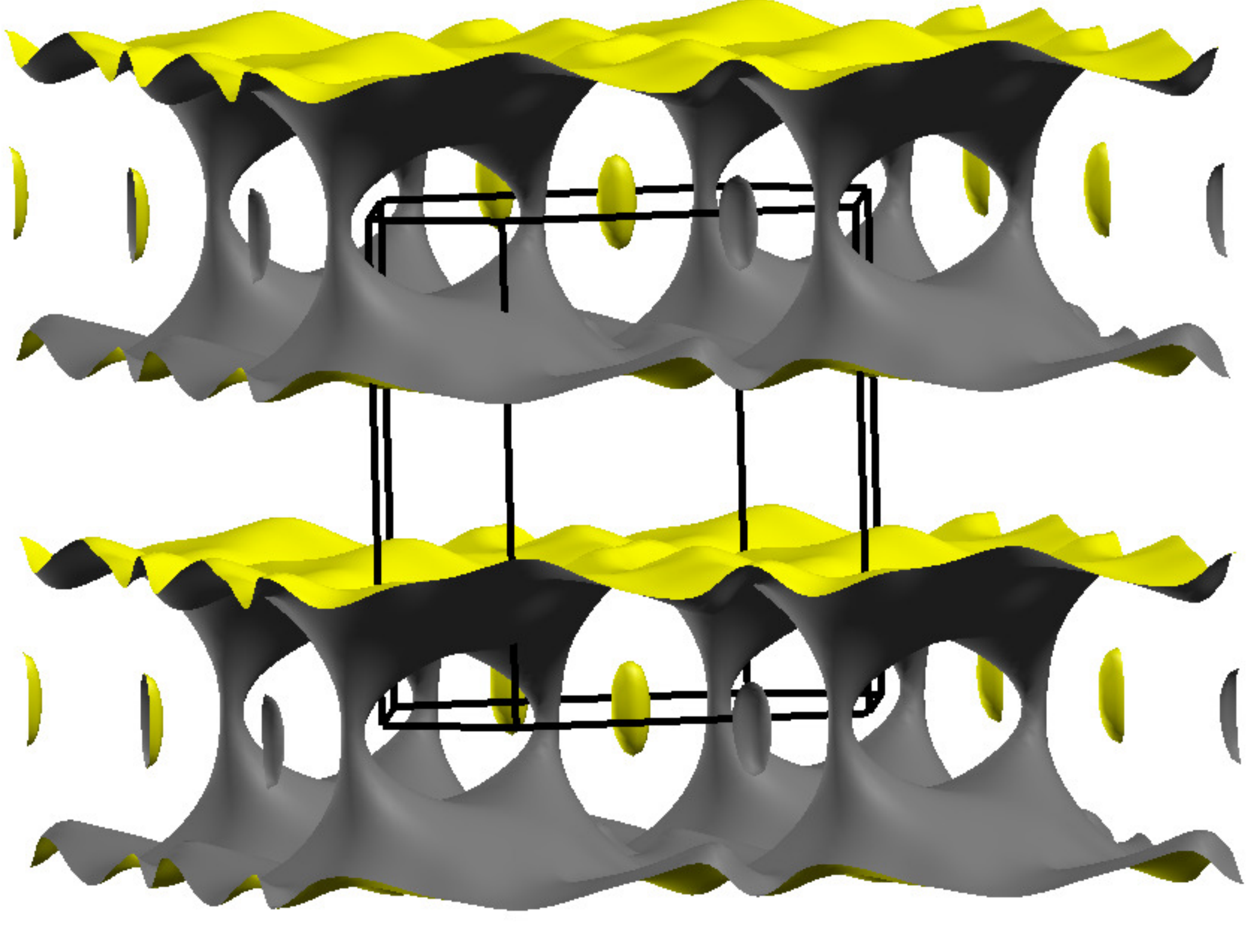}
\includegraphics[width=0.69\columnwidth]{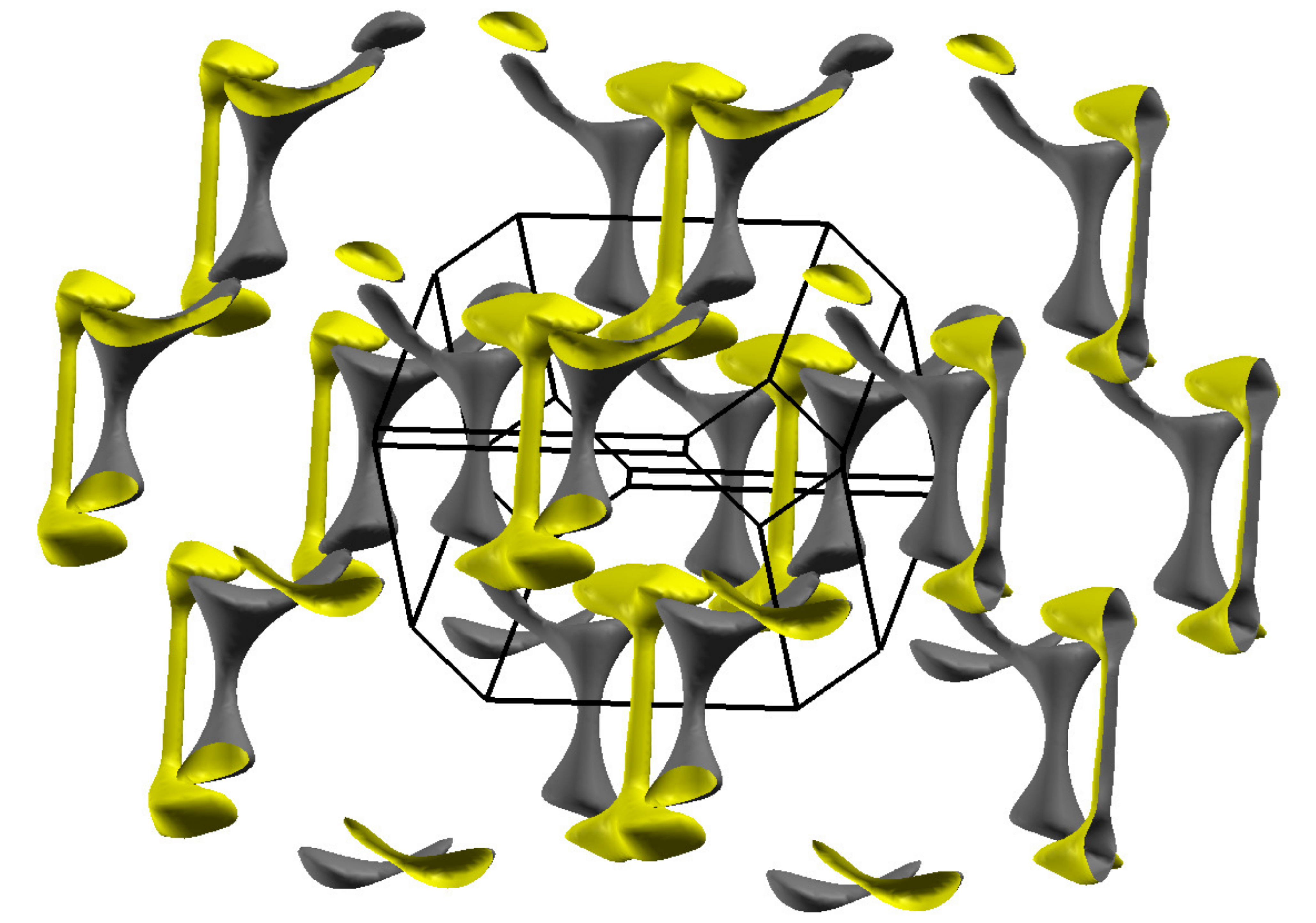}
\caption{Fermi surfaces of LuNiC$_2$ are displayed in a periodic zone scheme, 
for the sake of clarity without SOC, 
for the parent-type and the CDW modulated structure in the 
upper and lower panel, respectively.
Gray and yellow sides face to occupied and unoccupied states, respectively.}
\label{periodicscheme}
\end{center}
\end{figure}

The formation of a partial CDW gap is caused by the new Brillouin zone boundaries related to the
CDW structure modulation which doubles the real space periodicity along the orthorhombic 
$a$-axis.
The Brillouin zone boundaries of the CDW superstructure intersect with the 
pair of large Fermi surface sheets of parent-type structure of LuNiC$_2$ which 
forms an extended two-dimensional structure oriented perpendicular 
to the $b_1$ direction in the orthorhombic periodic zone scheme displayed in the upper
panel of Fig.~\ref{periodicscheme}. 
These quasi-planar (extended) Fermi surface sheets in the reciprocal space, manifest a 
quasi-one-dimensional electronic feature in real space. 
Our results of Fermi surface calculations of the CDW modulated structure displayed in 
a periodic zone scheme in the lower panel of Fig.~\ref{periodicscheme} reveal a
complete fragmentation of the extended Fermi surface of the orthorhombic parent structure 
into several isolated electron and hole pockets as the obvious consequence of introducing 
additional CDW super-zone boundaries in the context of a CDW gap formation. 

\section{Discussion}

The comparison of the electronic Sommerfeld coefficient obtained from the experimental 
specific heat data of LuNiC$_2$, $\gamma=0.83$(5)\,mJ/mol\,K$^2$
(see section~\ref{Cp_res}), with the bare Sommerfeld values, 
$\gamma_{\rm DFT}^{\rm CDW} \equiv N(E_{\rm F})\, k_{B}^{2} \pi^{2}/3= 0.90$ mJ/molK$^{2}$
for the CDW modulated structure and $\gamma_{\rm DFT}=2.43$ mJ/molK$^{2}$ for the parent structure
without CDW, reveals reasonably close agreement for the CDW model, but a rather large discrepancy for the 
DFT result of the orthorhombic parent structure of LuNiC$_2$. 
The latter is expect to be adopted at temperatures above $T_{\rm CDW}\simeq 450$\,K, 
at which electrical resistivity data indicate the partial CDW gap to vanish.  
On the other hand, there is a realistic agreement between the experimental Sommerfeld value of LuCoC$_2$, 
$\gamma=5.9$(1)\,mJ/mol\,K$^2$, and its corresponding calculated, bare electronic Sommerfeld value, 
$\gamma_{\rm DFT} = 3.82$ mJ/molK$^{2}$ of the CeNiC$_2$-type orthorhombic structure model of LuCoC$_2$. 
The approximate enhancement of the experimental value by a factor of 1.5 as compared to the calculated one 
relates to a weak to moderate electron-phonon effective mass enhancement, $\gamma = \gamma_{e} (1+\lambda_{ep})$, 
where $\lambda_{ep}\sim 0.5$ is the so called electron-phonon mass enhancement factor.

The quasi-one-dimensional metallic nature of LuNiC$_2$ at $T>T_{\rm CDW}$, 
which is indicated by the Fermi surface topology displayed in the upper panel of Fig.~\ref{periodicscheme},
is well corroborated by the significant anisotropy of the electrical resistivity at high temperatures, 
which, indeed, is lowest along the orthorhombic $a$-axis, i.e.\ perpendicular to the orientation of 
the quasi-planar Fermi surface sheets.
Within the CDW ordered state, i.e.\ below $T_{\rm CDW}\simeq 450$\,K, the partial CDW gap formation fragments 
these extended Fermi surface sheets into isolated Fermi surface pockets (compare Fig.~\ref{periodicscheme}) and, 
thus, conforms with the experimentally observed, more isotropic low temperature electrical resistivity 
as compared to the high-temperature state without CDW order.  
 
Some distinct differences between the observations from single crystal resistivity studies of LuNiC$_2$
in section~\ref{Cp_res} and correspondent results obtained for several
magnetic $R$NiC$_2$ compounds investigated by Shimomura {\sl et al.}~\cite{shimomura}
may relate to changes of the CDW modulations within the series of $R$NiC$_2$ compounds
(compare Ref.~\cite{PhysRevB.97.041103}).

\begin{figure}[t] 
\begin{center}
\includegraphics[width=0.9\columnwidth]{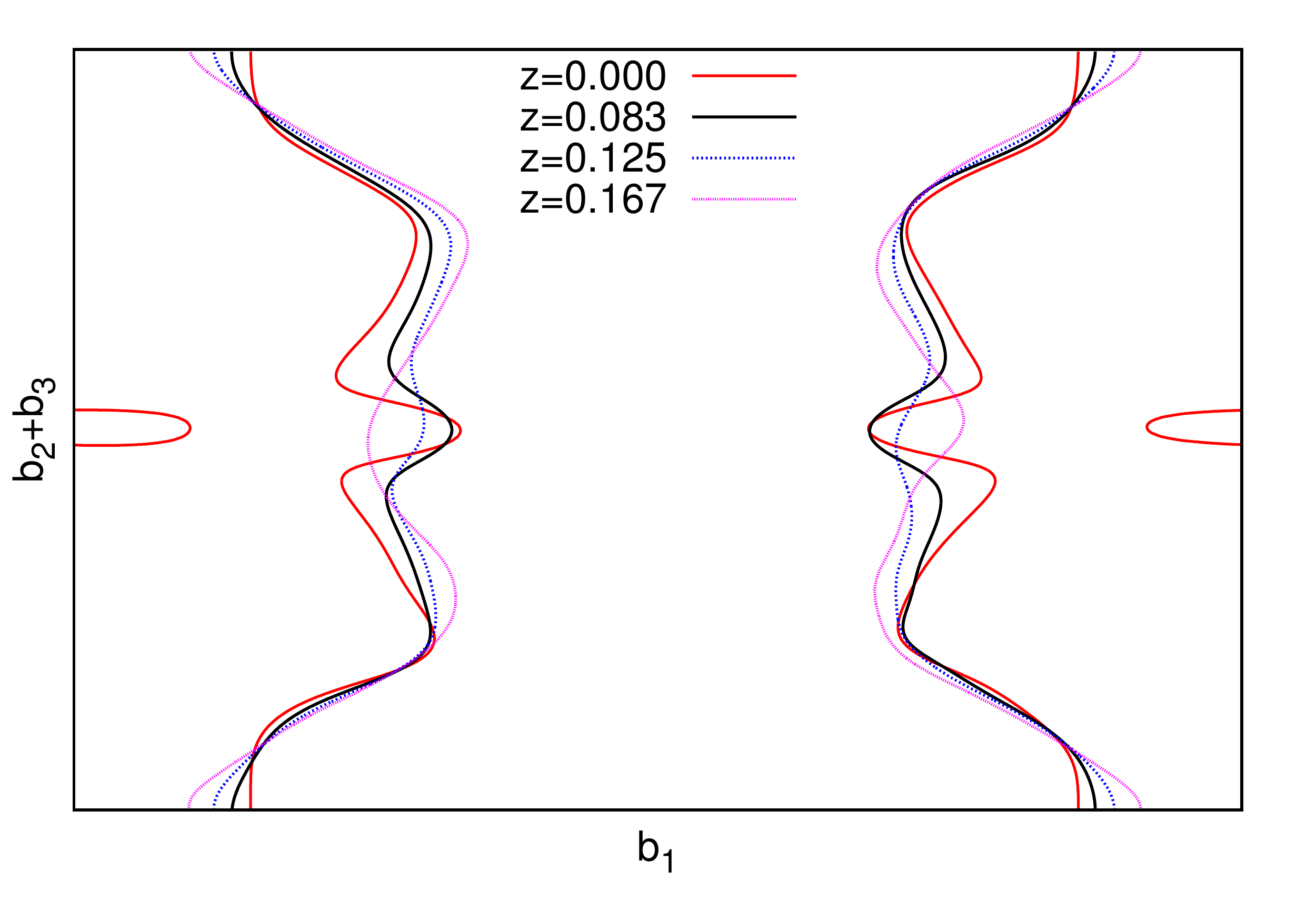}
\includegraphics[width=0.9\columnwidth]{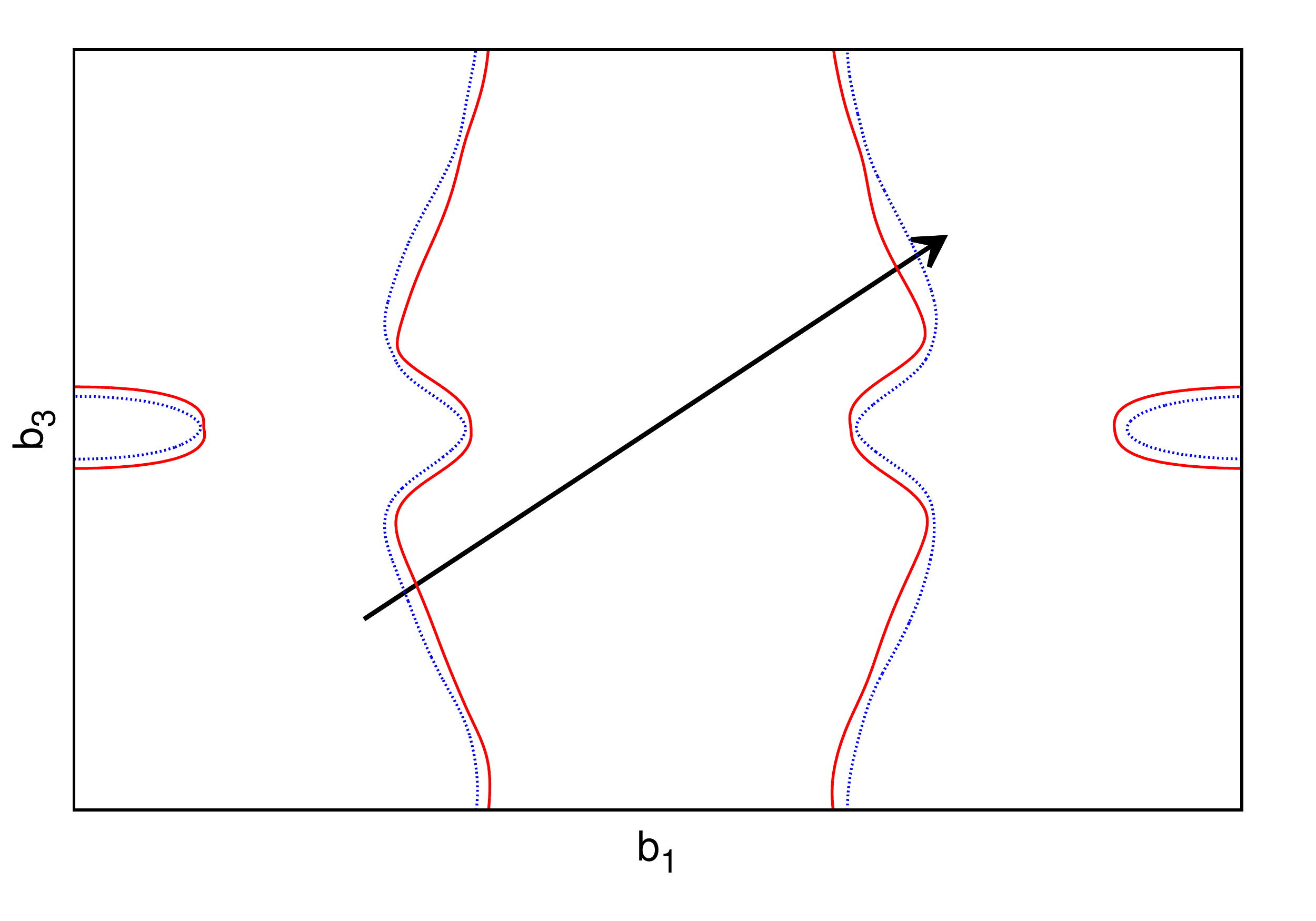}
\caption{Fermi surface contours of LuNiC$_2$ in parallel planes with
off-set $z$ away from the plane intersecting the $\Gamma$-point as labeled for 
planes span by reciprocal vectors $b_1$ and $b_2+b_3$ are displayed, 
for the sake of clarity without SOC, in the upper panel. 
The lower panel shows Fermi surface contours calculated with SOC in the plane span by $b_1$ and 
$b_3$ at $z=0$  (i.e.\ including the $\Gamma$-point) and the CDW structure modulation 
vector $\mathbf{Q}=(\frac1 2 , 0 , \frac1 2)$ is shown as a black arrow.
The two spin channels of SOC are indicated in the lower panel by red solid and blue dashed lines.}
\label{fs_cont}
\end{center}
\end{figure}

In order to discuss features of the Fermi surface of LuNiC$_2$ depicted in Fig.~\ref{fs} in closer detail,
we present in Fig.~\ref{fs_cont} slices of the Fermi surface as contour plots, where in the upper panel,
for the sake of clarity, SOC is neglected. 
The $z$-values therein refer to a parallel shift of these slices in units of the reciprocal lattice 
periodicity. Slices with identical orientation and position (except for $z=0$) as those shown in the upper panel 
of Fig.~\ref{fs_cont}, have earlier been presented by Laverock {\sl et al.}~\cite{laverock} for LaNiC$_2$ and SmNiC$_2$
(the basis vector setting with $b^{\star}$ in Ref.~\cite{laverock} corresponds to $b_2+b_3$ in the present work, while $a^{\star}=b_1$).
A direct comparison of the Fermi surface contours of LaNiC$_2$ and SmNiC$_2$ in Ref.~\cite{laverock}
with those of LuNiC$_2$ in Fig~\ref{fs_cont} shows that the latter is getting closer to the 
idealized Fermi-surface of a half-filled quasi-one-dimensional band with two (ideally planar) parallel sheets.
Overall, the Fermi surface of the CDW metal SmNiC$_2$ ($T_{\rm CDW}\simeq 148$\,K~\cite{murase}) reported by Laverock 
{\sl et al.}~\cite{laverock} appears significantly more corrugated and slightly more deviant from half-filling than that 
of LuNiC$_2$.

An important aspect is revealed from the slice of IRBZ which exactly hosts the 
experimentally observed wave vector $\mathbf{Q}=(\frac1 2 , 0 , \frac1 2)$ of CDW ordering,
namely the slice span by reciprocal lattice vectors $b_1$ and $b_3$ displayed in the lower panel of Fig.~\ref{fs_cont}.
The commensurate wave vector $\mathbf{Q}=(\frac1 2 , 0 , \frac1 2)$ is shown as black arrow and appears to
roughly conform with a Fermi surface nesting vector which connects two quasi-parallel parts of the Fermi 
surface contour, however, with a significant length mismatch of about 10\,\%. 
Moreover, various other nesting features of similar or even higher significance can be found for 
$q_n$-vectors which are clearly different from the structure modulation $\mathbf{Q}$.  
From the present results of Fermi surface calculations of LuNiC$_2$ in its orthorhombic parent-type structure
we can not confirm an obvious and dominant Fermi surface nesting at wave vector $\mathbf{Q}=(\frac1 2 , 0 , \frac1 2)$. 
In the case of SmNiC$_2$, Laverock {\sl et al.}~\cite{laverock} has indicated an even better match of the 
experimental structure modulation wave vector with a more extended Fermi surface nesting feature although
the CDW ordering temperature is only one third as compared to LuNiC$_2$.   

The absence of a clear correlation between the CDW ordering temperature and the degree of a Fermi surface nesting, 
which matches the structure modulation wave vector, in the series of $R$NiC$_2$ compounds seems in line with recent 
theoretical proposals by Johannes and Mazin~\cite{mazin2008} and by Gor'kov~\cite{gorkov2012} suggesting, that in 
real systems studied so far, CDW-formation is driven by $q$-dependent electron-phonon coupling rather than 
by a simple electronic Fermi surface nesting mechanism. Deeper experimental and theoretical investigations of 
LuNiC$_2$, e.g.\ as compared to YNiC$_2$, may be suited to clarify the key factors which govern CDW ordering
temperature in this system.

\section{Summary and Conclusions}

LuCoC$_2$ and LuNiC$_2$ have been prepared in single crystalline form. 
X-ray single crystal diffraction confirmed LuCoC$_2$ to 
crystallize in the non-centrosymmetric orthorhombic space group  $Amm2$
and, apart from weak extra reflections related to twinned superstructure modulations, 
also the room temperature single crystal XRD data of LuNiC$_2$ are rather well 
accounted for by this CeNiC$_2$-type structure model, however, with a significantly 
enhanced anisotropic displacement parameter $U_{11}$ of the Ni-site.
From the analyis of LuNiC$_2$ crystal data taken at 100\,K we derived
a model of a monoclinic superstructure (space group $Cm$) which suggests 
a commensurate CDW modulation of the orthorhombic parent structure. 
The largest deviation from $Amm2$ symmetry of the CeNiC$_2$-type structure model is observed 
for the Ni atoms which are displaced along the orthorhombic $a$-axis
in form of a Peierls-type distortion the Ni atom periodicity with pairwise shortened Ni---Ni distances 
of 3.208(2) \AA{} alternating with pairwise elongated distances of 3.682(2)~\AA.

Electrical resistivity and specific heat measurements of LuCoC$_2$ 
indicate a simple metallic behavior with neither CDW ordering nor superconductivity above 0.4\,K.
Analogous electrical resistivity studies of poly- as well as single crystalline samples of 
LuNiC$_2$, however, reveal a metallic state with a CDW transition at $T_{\rm CDW}\simeq 450$\,K
and a significant anisotropy of the electrical resistivity for $T>T_{\rm CDW}$. 
The high-temperature resistivity is lowest along the orthorhombic $a$-axis and highest along the $c$-axis. 
CDW ordering causes a reduction of anisotropy of the electrical resistivity at lower temperatures.  

The low temperature specific heat results reveal significantly different electronic Sommerfeld coefficients, 
$\gamma=5.9$(1)\,mJ/mol\,K$^2$ for LuCoC$_2$ and $\gamma=0.83(5)$\,mJ/mol\,K$^2$ for LuNiC$_2$, while DFT calculations 
of the state before CDW ordering suggests much closer values of the electronic density of states at the Fermi level, 
$N(E_{\rm F})=1.62$ states/eV\,f.u.\ for LuCoC$_2$ and $N(E_{\rm F})=1.03$ states/eV\,f.u.\ for LuNiC$_2$. 
The large reduction of the experimental Sommerfeld coefficient of LuNiC$_2$ as compared to the calculated, 
bare band structure value, $\gamma_{\rm DFT}=2.43$\,mJ/mol\,K$^2$, suggests that the Fermi surface of 
LuNiC$_2$ is strongly modified by CDW gap formation. DFT calculations based on the 
monoclinic CDW superstructure model, indeed, indicate the formation of a pronounced minimum
right at the Fermi level and the corresponding DFT result of the Sommerfeld coefficient of 
the CDW modulated state, $\gamma_{\rm DFT}^{\rm CDW}=0.90$\,mJ/mol\,K$^2$, 
reaches close agreement with the experimental value.

The fact that LuNiC$_2$ displays the highest CDW ordering temperature among isostructural $R$NiC$_2$ compounds,
even though a coincidence of the wave vector of the CDW modulation, $\mathbf{Q}=(\frac1 2 , 0 , \frac1 2)$,
with a predominant Fermi surface nesting vector remains weak, suggests a relevance of other mechanisms such as
CDW formation due to strongly $q$-dependent electron-phonon coupling proposed in Refs.~\cite{mazin2008,gorkov2012}.

\section*{Acknowledgements} 
Metallographic support by S. Stojanovic is gratefully acknowledged. 
We thank M. Abd-Elmeguid for fruitful discussions. 
The computational results have been achieved in part by using the Vienna Scientific Cluster (VSC)
within the project number $71074$, support of the VSC staff is gratefully acknowledged.


%


\end{document}